\documentclass[showpacs]{revtex4}

\usepackage{graphicx}
\usepackage{color}

\begin{document}

\title{Random matrix model for antiferromagnetism and
  superconductivity on a two-dimensional lattice}

\author{Beno\^\i t Vanderheyden} \affiliation{Department
  of Electrical Engineering and Computer Science and SUPRATECS, \\ 
  Universit{\'e} de Li\`ege, B28\\ B-4000 Li\`ege
  (Sart-Tilman), Belgium}

\author{A. D. Jackson} \affiliation{The Niels Bohr International Academy, The Niels Bohr Institute,
Blegdamsvej 17, DK-2100 Copenhagen \O, Denmark}

\date{\today}

\begin{abstract}

We suggest a new mean-field method for studying the thermodynamic
competition between magnetic and superconducting phases in a
two-dimensional square lattice. A partition function is constructed by
writing microscopic interactions that describe the exchange of density
and spin fluctuations.  A block structure dictated by spin, time-reversal,
and bipartite symmetries is imposed on the single-particle
Hamiltonian.  The detailed dynamics of the interactions are neglected
and replaced by a normal distribution of random matrix elements. The
resulting partition function can be calculated exactly. The
thermodynamic potential has a structure which depends only on the
spectrum of quasiparticles propagating in fixed condensation
fields, with coupling constants that can be related directly to the
variances of the microscopic processes. The resulting phase diagram
reveals a fixed number of phase topologies whose realizations depend
on a single coupling parameter ratio, $\alpha$. Most phase topologies
are realized for a broad range of values of $\alpha$ and can thus be
considered robust with respect to moderate variations in the detailed
description of the underlying interactions.

\end{abstract}


\maketitle

\section{Introduction}

Studies of high-temperature superconductors have revealed a rich phase
diagram, with coexisting magnetic and superconducting correlations.
These phase structures can be complex and include $d$-wave pairing,
stripes, or the pseudogap phenomenon.\cite{Dagotto94, Scalapino06,
  Lee2006, Carlson2008} Theoretical models and numerical studies on a
lattice indicate that the richness of the phase structure results from
a delicate energy balance between competing states.\cite{Scalapino06,
  Yao2007} It follows that model predictions can be sensitive to small
changes in the parameters of the theory or to details of the numerical
approach. The question then arises of which properties of the phase
diagram are constrained by the basic underlying symmetries and which
are sensitive to the detailed dynamics of the interactions and to
numerical approximations.

The purpose of this paper is to address this question with a new
mean-field approach. The method is based on random matrix theory and
consists in constructing a Hamiltonian that retains the basic spin,
time-reversal, and bipartite symmetries of the problem but simplifies
the dynamics of the interactions considerably. The theory is radically
different from the familiar Hubbard or $t$-$J$ model. Here, we
construct the model at a deeper microscopic level and describe
interactions that are mediated by the exchange of density and spin
fluctuations. This construction is inspired by random matrix models of
the strong interaction, for which the QCD interactions are mediated by
single-gluon exchange. Although natural in QCD, a microscopic
description involving bosonic fields may be more controversial in the
context of high-$T_c$ superconductors. Such an approach is similar to
low-energy effective theories of antiferromagnets and
superconductors\cite{Kampfer2005,Brugger2006,Brugger2006a,Brugger2007a,
  Brugger2007} or to the antiferromagnetic spin-fluctuation exchange
theory.\cite{Bickers1989, Moriya1991, Monthoux1999, Chubukov2002}
Here, in contrast to these models, no particular assumption is made
regarding the detailed dynamics of the exchange fields. Instead, we
adopt a coarse description in which the block structure of the
interaction matrix is dictated by the underlying symmetries of the
Hamiltonian while individual matrix elements are drawn at random.

A random matrix approach offers three advantages. First, since the
theory is constructed at a more microscopic level, it allows us to
relate global properties of the phase diagram to specific microscopic
mechanisms. Second, the simplified dynamics produces a mean-field
model that can be solved exactly: the gap equations are polynomial.
Their roots can therefore be studied as a function of the coupling
parameters of the theory. Third, in the vicinity of critical points,
the thermodynamic potential has a Landau-Ginzburg form in which the
expansion coefficients satisfy specific symmetry constraints inherited
from the deeper microscopic level. These constraints help us to
identify those topologies that can be realized in the system and rule
out those that violate the constraints.

The motivation for applying methods used in QCD to the high-$T_c$
problem follows from the strong analogies existing between these
systems.  First, the restoration of chiral symmetry with increasing
quark density can be understood from the analogous behaviors of QCD and
metamagnets. The chiral condensate plays the role of a staggered
magnetization which vanishes abruptly as an external magnetic field is
increased, driving the system through a first-order phase
transition.\cite{Halasz1998} Second, single-gluon exchange is
attractive in the antitriplet channel and leads to the Cooper pairing
of quarks. This form of pairing can lead to a long-range order called
\emph{color superconductivity}.\cite{RajagopalWilczek00, Alford2001,
  Alford2007} The degrees of freedom that are involved in pairing are
different from those involved in the chiral broken phase of QCD, so
that color superconductivity competes directly with the breaking of
chiral symmetry.

Random matrix methods have been extensively applied for studying the
phase diagram of
QCD.\cite{JacksonVerbaarschot96,HalaszJacksonVerbaarschot96,%
  Halasz1998,KleinToublanVerbaarschot03,KleinToublanVerbaarschot05}
In recent works,\cite{VanJac00,VanJac00b,VanJac03,VanJac05} we
studied the phase diagram of QCD with three colors and two flavors as
a function of temperature and quark chemical potential.  The partition
function was  constructed as an integral over random matrices
that mimicked the basic structure of quark-quark interactions but neglected
their detailed dynamics. These matrices were given a block structure
that reflected the spin, color, and flavor symmetries of one-gluon
exchange. Inside a given block matrix, no further correlations were
assumed among the matrix elements, which were then drawn at random on
a normal distribution.  This approach produced a mean-field model that
could be solved exactly. The resulting effective potential gave
polynomial gap equations whose roots could be determined analytically
or numerically. The effective potential contained a single free
parameter, defined as a coupling-constant ratio that measured the
relative strength of the interactions in the chiral and diquark
channels. As this coupling ratio was varied, the phase structure
passed through a restricted number of distinct topologies. Moreover,
starting with coupling constants with the values appropriate for
single-gluon exchange, large variations were required to alter the
topology of the phase diagram. We thus concluded that the QCD phase
diagram was robust against moderate variations in the detailed
dynamics of the interactions.

In general, random matrix models are useful in providing a global
picture of the phase diagram. Being mean-field in nature, such a
picture is only a starting point that requires the proper inclusion of
thermal, quantum, and spatial fluctuations (which play an important
role in the high-$T_c$ problem as a consequence of the Mermin-Wagner
theorem\cite{Mermin1966}) if it is to be quantitatively reliable.
Nevertheless, the random matrix approach can be useful in providing an
overview of the strength of the order parameters and their sensitivity
to coupling parameters and can thus serve as a means for identifying
those characteristics of the phase diagram that are protected by
symmetry.

It is worth noting that the ``random'' part of the theory has nothing
to do with disorder. Instead, the philosophy here consists in
constructing a Hamiltonian with a block structure dictated by the
underlying symmetries of the problem and replacing individual matrix
elements by random variables, conventionally drawn on a normal
distribution.  This construction can be regarded as equivalent to
integrating over many Hamiltonians that meet fundamental symmetry
requirements but which differ from one another in the detailed
implementation of the dynamics of the interaction.  The theory can
thus appear elaborate at first sight since it involves a large number
of statistical variances. However, the number of free parameters
decreases at each step of the calculations so that the theory becomes
simpler as one proceeds towards the solution of the problem. In fact,
the final form of the thermodynamic potential depends only on a single
parameter ratio, and its functional form has a well-defined structure
that can be interpreted in terms of quasiparticle energies. Hence,
most of the effort involved in constructing the model is ``upfront''
but worthwhile since it provides relationships between the global
phase diagram and the microscopic parameters of the theory. Overall,
the procedure is relatively simple and could be implemented in many
other problems.

We will consider a fermion system on a two-dimensional square lattice
and construct its partition function at finite temperature and finite
chemical potential. In Sec.~\ref{s:RM}, we show that extension of
the methods used in QCD poses some challenges.  First, while the basic
interactions of QCD are naturally formulated in terms of quarks
exchanging gluons, no such natural description is available at an
elementary level for the cuprates. We will thus explicitly assume that
the interactions can be described as the exchange of density and spin
fluctuations.  The structure of the interaction between the fermions
and the fluctuation fields is dictated by $SU(2)$-spin, time-reversal,
and bipartite symmetries. We will show that this formulation leads to
a four-fermion effective potential whose terms can be compared to
those of the Hubbard model.  A second challenge is the need to account
for the $d$-wave character of the superconducting order parameter,
which forces us to introduce an explicit momentum dependence of the
fermion states. As we are seeking to construct a model that suppresses
as many details of the interaction as possible, we limit ourselves to
a ``coarse-grained'' momentum description in which the first Brillouin
zone is divided into four sectors in order to mimic the symmetry
patterns of the antiferromagnetic and superconducting order parameters.

Section~\ref{s:TP} is devoted to the derivation of the effective
potential. We find that it is possible to construct a theory in which
antiferromagnetism and $d$-wave superconductivity are favored and
compete, while the $s$-wave channel is repulsive. Such a theory
gives greater statistical weight to spin-fluctuation fields with a
large momentum exchange. When deriving the effective potential, we
will see that the theory simplifies considerably as one proceeds
through the calculations. The initial model of Sec.~\ref{s:RM} 
involves as many as eight different variance parameters; the final 
thermodynamic potential depends on a single parameter ratio.  
Its interpretation in terms of quasiparticle energies is also
given in Sec.~\ref{s:TP}.

We study the resulting phase diagrams in Sec.~\ref{s:PD}, where we
show that, as was the case for QCD, there are only a finite number
of possible topologies. As in the QCD problem, the topology of the phase
structure changes gradually as the ratio of coupling constants is
varied. Section~\ref{s:C} contains a summary of our main findings and
our conclusions.

\section{Construction of the random matrix model}
\label{s:RM}

We consider a system of electrons on a two-dimensional square lattice
and model the competition between magnetic and superconducting orders
as a function of the chemical potential, $\mu$, and the temperature,
$T$.  We construct interactions that satisfy spin, time-reversal, and
bipartite symmetries with simplified but integrable dynamics. This
approach ensures that the properties of the model arise solely as a
consequence of the symmetries.

\subsection{Order parameters and the parameterization of momenta and frequencies}
\label{ss:OP-and-momentum}

We wish to define random matrix correlators that mimic the basic
structure of antiferromagnetism and superconducting order
parameters. Working at finite temperature $T$ in an imaginary time
formalism, the antiferromagnetic order parameter assumes the form
\begin{eqnarray}
  \mathbf{m}_{AF} = \left\langle \sum_{\mathbf{p}\,\omega_{n}\alpha\beta} 
\psi^{\dagger}_\alpha(\mathbf{p+Q},\omega_n)\,
\mbox{\boldmath$\sigma$}_{\alpha\beta}^{\phantom{\dagger}}\, 
\psi^{\phantom{\dagger}}_\beta(\mathbf{p},\omega_n)\right\rangle,
\label{AF-OP-microscopic}
\end{eqnarray}
where $\mathbf{p}$ are momenta in the first Brillouin zone,
$\mathbf{Q}= (\pi \hbar/a,\pi \hbar/a)$ is the AF ordering vector ($a$
is the lattice spacing), $\omega_n = (2 n +1 ) \pi T$ are fermion
Matsubara frequencies, $\alpha$ and $\beta$ are spin indices,
{\boldmath $\sigma$} are the spin Pauli matrices, and $\langle \ldots
\rangle = \mathrm{Tr}(\ldots e^{-\beta H})$ denotes a thermal
average. Similarly, the $d$-wave order parameter is given as
\begin{eqnarray} 
m_{SC-d} = \left\langle \sum_{\mathbf{p},\omega_n} g(\mathbf{p})\,
\psi_\uparrow
(\mathbf{p},\omega_n)\,\psi_\downarrow(-\mathbf{p},-\omega_n)\right\rangle,
\label{SC-d-OP-microscopic}
\end{eqnarray}
where
\begin{eqnarray}  
g(\mathbf{p}) = \cos\left({p_x a \over \hbar}\right) - \cos\left({p_y
  a \over \hbar} \right)
\label{gp-d-wave}
\end{eqnarray}
is the $d$-wave form factor. 

Our aim is to construct correlators that mimic the momentum
couplings in Eqs.~(\ref{AF-OP-microscopic}) and
(\ref{SC-d-OP-microscopic}) on a coarse level.  First, we divide the
Brillouin zone into four regions related to one another by
either a momentum shift in $\mathbf{Q}$,
\begin{eqnarray}
  \{\mathbf{p}\} \mapsto Q \{\mathbf{p}\} = \{\mathbf{p} + \mathbf{Q}\}.
\end{eqnarray}
or by momentum reversal,
\begin{eqnarray}
  \{\mathbf{p}\} \mapsto P \{\mathbf{p}\} =  - \{\mathbf{p}\}.
\end{eqnarray}
Next, we replace the exact form factor $g$ in Eq.~(\ref{SC-d-OP-microscopic}) 
by the simplified form factor
\begin{eqnarray}
  \phi_d(\mathbf{p}) = \mathrm{sign}\left(g(\mathbf{p})\right),
\label{simplified-form-factor-d}
\end{eqnarray}
where $\phi_d$ is a crude approximation of $g$.  It neglects its variation with
momentum, which is related to the detailed shape of the wave function,
but exhibits the same $d$-wave symmetry as $g$, changing sign for
every ninety degree rotation in the Brillouin zone.  We believe that
such an approximation captures the essential symmetry of the problem
and is sufficient to describe $d$-wave pairing.  

\begin{figure}[ht]
\begin{center}
\begin{picture}(0,0)%
\includegraphics{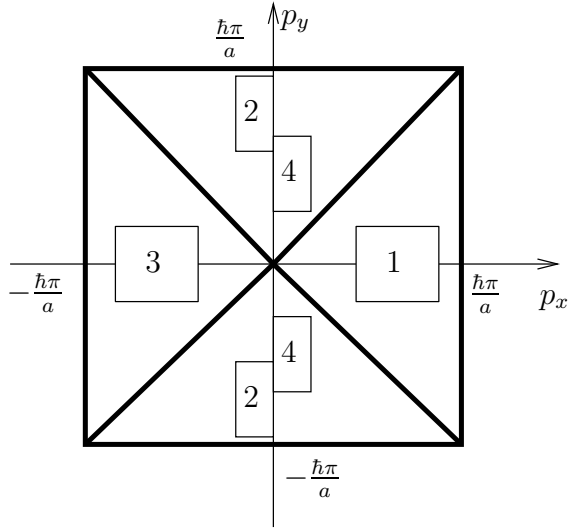}%
\end{picture}%
\setlength{\unitlength}{4144sp}%
\begingroup\makeatletter\ifx\SetFigFontNFSS\undefined%
\gdef\SetFigFontNFSS#1#2#3#4#5{%
  \reset@font\fontsize{#1}{#2pt}%
  \fontfamily{#3}\fontseries{#4}\fontshape{#5}%
  \selectfont}%
\fi\endgroup%
\begin{picture}(3340,3210)(415,-3718)
\put(2071,-691){\makebox(0,0)[lb]{\smash{{\SetFigFontNFSS{12}{14.4}{\familydefault}{\mddefault}{\updefault}{\color[rgb]{0,0,0}$p_y$}%
}}}}
\put(3196,-2356){\makebox(0,0)[lb]{\smash{{\SetFigFontNFSS{12}{14.4}{\familydefault}{\mddefault}{\updefault}{\color[rgb]{0,0,0}${\hbar \pi \over a}$}%
}}}}
\put(1666,-826){\makebox(0,0)[lb]{\smash{{\SetFigFontNFSS{12}{14.4}{\familydefault}{\mddefault}{\updefault}{\color[rgb]{0,0,0}${\hbar \pi \over a}$}%
}}}}
\put(3618,-2367){\makebox(0,0)[lb]{\smash{{\SetFigFontNFSS{12}{14.4}{\familydefault}{\mddefault}{\updefault}{\color[rgb]{0,0,0}$p_x$}%
}}}}
\put(430,-2351){\makebox(0,0)[lb]{\smash{{\SetFigFontNFSS{12}{14.4}{\familydefault}{\mddefault}{\updefault}{\color[rgb]{0,0,0}$-{\hbar \pi \over a}$}%
}}}}
\put(2086,-3460){\makebox(0,0)[lb]{\smash{{\SetFigFontNFSS{12}{14.4}{\familydefault}{\mddefault}{\updefault}{\color[rgb]{0,0,0}$-{\hbar \pi \over a}$}%
}}}}
\put(2701,-2176){\makebox(0,0)[lb]{\smash{{\SetFigFontNFSS{12}{14.4}{\familydefault}{\mddefault}{\updefault}{\color[rgb]{0,0,0}$1$}%
}}}}
\put(1846,-1276){\makebox(0,0)[lb]{\smash{{\SetFigFontNFSS{12}{14.4}{\familydefault}{\mddefault}{\updefault}{\color[rgb]{0,0,0}$2$}%
}}}}
\put(1846,-2986){\makebox(0,0)[lb]{\smash{{\SetFigFontNFSS{12}{14.4}{\familydefault}{\mddefault}{\updefault}{\color[rgb]{0,0,0}$2$}%
}}}}
\put(1261,-2176){\makebox(0,0)[lb]{\smash{{\SetFigFontNFSS{12}{14.4}{\familydefault}{\mddefault}{\updefault}{\color[rgb]{0,0,0}$3$}%
}}}}
\put(2071,-2716){\makebox(0,0)[lb]{\smash{{\SetFigFontNFSS{12}{14.4}{\familydefault}{\mddefault}{\updefault}{\color[rgb]{0,0,0}$4$}%
}}}}
\put(2071,-1636){\makebox(0,0)[lb]{\smash{{\SetFigFontNFSS{12}{14.4}{\familydefault}{\mddefault}{\updefault}{\color[rgb]{0,0,0}$4$}%
}}}}
\end{picture}
 
 \caption{Parametrization of the first Brillouin zone.  Regions $1$
    and $2$ are related by a momentum shift by $\pm {\mathbf{Q}}=\pm (\pi
    \hbar/a, \pi \hbar/a) $, as are regions $3$ and $4$. Regions $1$
    and $3$ and regions $2$ and $4$ are related by momentum reversal.}
  \label{f:BZ}
\end{center}
\end{figure}

The particular form for $\phi_d$ guides our parametrization of
momentum states.  Four momentum regions are chosen as the sectors in
which $\phi_d$ has a given sign.  One possible division of the
Brillouin zone is shown in Fig.~\ref{f:BZ}.  The approximate form
factor $\phi_d$ is $-1$ in regions $1$ and $3$ and $+1$ in regions $2$
and $4$.  States in regions $1$ and $3$ are related to those in
regions $2$ and $4$ by a shift of $\pm \mathbf{Q}$.  Regions $1$ and
$2$ are related to regions $3$ and $4$ by momentum reversal. Inside
each region, states are labeled by an index $i = 1, \ldots, M$, where
$M$ scales with the total number of lattice sites and $M\to \infty$ in
the thermodynamical limit. These states also count the different
Matsubara frequencies. From one region to another, the states are
parametrized as follows. If a given index $i$ refers to a state
$(\mathbf{p},\omega_n)$ in region $1$, then the states labeled with
the index $i$ in $2$, $3$, and $4$, respectively, correspond to
$(\mathbf{p}+\mathbf{Q},\omega_n)$, $(-\mathbf{p},-\omega_n)$, and
$(-\mathbf{p}-\mathbf{Q},-\omega_n)$.  (Note the change of sign in the
frequencies of the last two terms.)

With this parametrization, the AF order parameter is written as 
\begin{eqnarray}
  \mathbf{m}_{AF}=\left\langle \sum_{r,s=1}^4\sum_{i,j=1}^{M} 
\sum_{\alpha,\beta=\uparrow,\downarrow} 
\psi^{\dagger}_{r,i,\alpha}
 \mbox{\boldmath$\sigma$}_{\alpha\beta} \,\left(\Gamma_{AF}\right)_{r,s}\,\delta_{ij}\,
\psi^{\phantom{\dagger}}_{s,j,\beta}
\right\rangle,
\label{mAF-RM}
\end{eqnarray}
where $r$ and $s$ are region indices and the four-by-four matrix
$\Gamma_{AF}$ couples momenta separated by $\mathbf{Q}$:
\begin{eqnarray}
\Gamma_{AF}= (\sigma_1)_Q \otimes (\mathbf{1})_P =
\left(
\begin{array}{cccc}
0 & 1 & 0 & 0\\
1 & 0 & 0 & 0\\
0 & 0 & 0 & 1\\
0 & 0 & 1 & 0\\
\end{array}
\right).
\label{GammaAF}
\end{eqnarray}
Similarly, the superconducting order parameter has the form
\begin{eqnarray}
  m_{SC-d} = \left\langle \sum_{r,s=1}^4\sum_{i,j=1}^{M} 
 \psi_{r,i,\uparrow} \left(\Gamma_{SC-d}\right)_{r,s} \,\delta_{ij}\,\psi_{s,j,\downarrow}
 \right\rangle
\label{mSC-d-RM}
\end{eqnarray}
where $\Gamma_{SC-d}$ is now blind to shifts by $\mathbf{Q}$ and couples
states with opposite momenta with a sign dictated by the $d-$wave form
factor of Eq.~(\ref{simplified-form-factor-d}):
\begin{eqnarray}
  \Gamma_{SC-d} = (\phi_d(\mathbf{p}))_Q \otimes (\sigma_1)_P =
(-\sigma_3)_Q \otimes (\sigma_1)_P = 
\left(
\begin{array}{cccc}
0 & 0 & -1 & 0\\
0 & 0 & 0 & 1\\
-1 & 0 & 0 & 0\\
0 &  1 & 0 & 0\\
\end{array}
\right).
\label{GammaSC-d}
\end{eqnarray}
Note that with our parametrization, the Kronecker symbol $\delta_{ij}$
in Eqs.~(\ref{mAF-RM}) and (\ref{mSC-d-RM}) automatically
selects equal Matsubara frequencies for the AF order parameter
and opposite frequencies for the superconducting SC order parameter.

Similar arguments can be followed for an $s-$wave order
parameter. With an isotropic form factor $\phi_s(\mathbf{p}) = 1$, the
corresponding momentum projector is given as
\begin{eqnarray}
  \Gamma_{SC-s} = (\mathbf{1})_Q\otimes(\sigma_1)_P = 
\left(
\begin{array}{cccc}
0 & 0 & 1 & 0\\
0 & 0 & 0 & 1\\
1 & 0 & 0 & 0\\
0 &  1 & 0 & 0\\
\end{array}
\right),
\label{GammaSC-s}
\end{eqnarray}
and the $s$-wave order parameter has the form given by Eq.~(\ref{mSC-d-RM})
with $\Gamma_{SC-d}$ replaced by $\Gamma_{SC-s}$. 

\subsection{Constraints imposed by symmetries}
\label{ss:basic-symmetries}

We now turn to the construction of the random matrix
interactions. Inspired by random matrix models for QCD\cite{VanJac00,
  VanJac00b}, we write the partition function of the system as a path
integral
\begin{eqnarray}
  Z(\mu,T) = \int {{\cal D} \psi^\dagger}\,{{\cal D}
    \psi^{\phantom{\dagger}}}\, d
  H_{\mathrm{int}}\;P(H_\mathrm{int})\;e^{- \psi^\dagger (H_0 + H_{\mathrm{int}}) \psi^{\phantom{\dagger}}},
\label{partition-function}
\end{eqnarray}
where $\psi^\dagger$ and $\psi^{\phantom{\dagger}}$ are independent
fermion fields, $H_\mathrm{int}$ is a matrix describing the random
interaction with elements distributed according to the distribution 
$P(H_{\mathrm int})$ to be defined below, and $H_0$ is the non-interacting 
part of the single-particle Hamiltonian which contains temperature and 
chemical-potential terms.

This formulation is radically different from the more familiar Hubbard
or $t$-$J$ models, where the interaction terms are described by
effective four-fermion potentials. Here, instead, the interactions are
described at a deeper microscopic level so that fermions interact 
with fluctuation fields via current terms, 
$\psi^\dagger H_{\mathrm{int}}\psi$.  This formulation 
is directly inspired by the Yang-Mills
Lagrangian for the strong interaction mediated by gluon exchange.  In
the context of the high-$T_c$ problem, these fluctuation fields can
describe either interactions carried by phonons, antiferromagnetic
fluctuations, or more complex effective interactions. Here, we will
not attempt to identify the nature of these fields nor to specify their 
dynamics. We will assume only that interactions can be described in 
terms of such fields and consider how their description is constrained 
by the symmetries of the system.

{\bf Interaction terms}. According to the parametrization introduced
in Sec.~\ref{ss:OP-and-momentum}, the fermion fields are described by
spinors with eight components (four momentum regions and two spins).
Similarly, the random matrices $H_{\mathrm{int}}$ are composed of
$8\times 8$ block matrices of size $M\times M$. Due to the symmetries
of the system, these blocks are not completely independent.

Consider first the constraints imposed by $SU(2)$. For a fixed pair of
momentum region indices, e.g., $(r,s)$, we can write
\begin{eqnarray}
  \left(\begin{array}{cc} H_{\mathrm{int}, r,s,\uparrow,\uparrow} &
    H_{\mathrm{int},r,s,\uparrow,\downarrow}\\ H_{\mathrm{int},r,s,\downarrow,\uparrow}
    & H_{\mathrm{int},r,s,\downarrow,\downarrow}\\
\end{array}\right)
 = \sum_{\mu=0}^3 \sigma_{\mu}\,H_{\mu ; r,s},
\label{Ars-decomposed}
\end{eqnarray}
where $\sigma_\mu = (\mathbf{1},\mbox{\boldmath $\sigma$})$,
$H_{\mu;r,s}$ with $\mu=0$ represents density-fluctuation fields, and
$H_{\mu;r,s}$ with $\mu=1,2,3$ describes spin-fluctuation fields. The
interaction Hamiltonian can be made invariant under a spin unitary
transformation, $U$, by requiring that the vector
$(H_{1;r,s},H_{2;r,s},H_{3;r,s})$ simultaneously undergoes a space
rotation $R$, with $R$ satisfying $U^{\dag}\sigma_i U = R_{ij}
\sigma_j$. The partition function itself,
Eq.~(\ref{partition-function}), is then made invariant under $SU(2)$
transformations by requiring that the probability distribution,
$P(H_\mathrm{int})$, is invariant under the corresponding spatial
rotations $R$.

Consider next constraints related to time-reversal
invariance. Following the work of Monthoux on spin-fluctuation
exchanges\cite{Monthoux1999}, the fields $H_\mu$ are taken to be real
in coordinate representation --- they do not carry an electric
charge. As will be seen below, the integration over these fields
produces a four-fermion interaction that contains squares of density
terms, $\sim (\psi^\dagger\psi)^2$, and spin currents, $\sim
(\psi^\dagger\mbox{\boldmath $\sigma$}\psi)^2$, thus leading to a
two-body potential with a time-reversal symmetry. Now, because the
fluctuation fields $H_\mu$ are real, the Fourier components of their
matrix elements must satisfy the constraint
\begin{eqnarray}
  \left(\mathbf{p},\omega_n|H_\mu|\mathbf{q},\omega_m\right)
  = \left(-\mathbf{p},-\omega_n|H_\mu|-\mathbf{q},-\omega_m\right)^*.
\label{reality-momentum-representation}
\end{eqnarray}
If one divides the Brillouin zone into two subspaces of states
$\{\mathbf{p},\omega_n\}$ (regions $1$ and $2$) and $\{-\mathbf{p},-\omega_n\}$ (regions
$3$ and $4$) and adopt the parametrization introduced in
Sec.~\ref{ss:OP-and-momentum}, this condition can be cast in the form
\begin{eqnarray}
  H_\mu = P H^*_\mu P,
\end{eqnarray}
where
\begin{eqnarray}
  P = \left(\begin{array}{cc}0 & 1\\1 & 0\end{array}\right)
\end{eqnarray}
reverses both momentum and frequency.  Hence, the matrices $H_\mu$
must have the block-structure
\begin{eqnarray}
  H_\mu = \left(\begin{array}{cc}B_\mu & C_\mu\\C^*_\mu &
    B^*_\mu\end{array}\right),
\label{block-structure-T-reversal}
\end{eqnarray}
where $B_\mu$ are Hermitian and $C_\mu$ are complex
symmetric.\footnote{Expressed here in the momentum basis, the matrix
  $H_{\mu}$ of size $N\times N$ contains $N^2/4$ independent elements
    for $B_\mu$ and $N^2/4+N/2$ independent elements for $C_\mu$,
    hence a total of $N(N+1)/2$ independent elements. This is also the
    number of independent elements found in the more familiar
    coordinate representation, where $H_\mu$ is
    real-symmetric.}

Finally, we turn to the bipartite symmetry, appropriate for a square
lattice, composed of two interpenetrating sublattices $A$ and
$B$. Consider the transformation
\begin{eqnarray}
  \psi(\mathbf{r}) \mapsto \left\{\begin{array}{ccc}
                     + \psi(\mathbf{r}) & \quad & \textrm{if~}\mathbf{r}\in A,\\
                     - \psi(\mathbf{r}) & \quad & \textrm{if~}\mathbf{r}\in B.\\
                 \end{array}\right.
\label{bipartite-coordinate}
\end{eqnarray}
The full Hamiltonian $H$ is not expected to be invariant under such a
transformation because the kinetic terms couple fields defined on
neighboring sites. However, we will assume that the \emph{interaction}
part of the Hamiltonian, $H_{\mathrm{int}}$, is bipartite invariant
(as is the case for the $U$ term in the Hubbard model). We first 
determine the momentum representation of Eq.~(\ref{bipartite-coordinate}).
Using the definition of $\mathbf{Q}=(\pi,\pi)\,\hbar/a$,
Eq.~(\ref{bipartite-coordinate}) can be rewritten as $\psi(\mathbf{r})
\mapsto \exp(i
\mathbf{Q}\cdot\mathbf{r}/\hbar)\,\psi(\mathbf{r})$. Then, in the
momentum representation, the bipartite transformation takes the form
\begin{eqnarray}
  \psi(\mathbf{p}) \mapsto  Q \,\psi(\mathbf{p}) = \psi(\mathbf{p}+\mathbf{Q}).
\end{eqnarray}
Dividing the Brillouin zone in the two subspaces $\{\mathbf{p}\}$
(regions $1$ and $3$) and $\{\mathbf{p}+\mathbf{Q}\}$ (regions $2$ and
$4$), the bipartite invariance of $H_\mu$ is written as
\begin{eqnarray}
  H_{\mu} = Q H_{\mu} Q,
\end{eqnarray}
with 
\begin{eqnarray}
  Q = \left(\begin{array}{cc}1 & 0\\0 & 1\\\end{array}\right).
\end{eqnarray}
$H_{\mu}$  must then have the block-structure
\begin{eqnarray}
  H_{\mu} = \left(
                      \begin{array}{cc}
                        D_\mu & E_\mu \\
                        E_\mu & D_\mu \\
                      \end{array}
                    \right),
\label{bipartite-block-structure}
\end{eqnarray}
where $D_\mu$ and $E_\mu$ are Hermitian. An alternative, but weaker,
requirement on $H_\mu$ could be that matrix elements between states
with momenta $\mathbf{p_1}+\mathbf{Q}$ and $\mathbf{p_2}+\mathbf{Q}$
are equal to those between $\mathbf{p_1}$ and $\mathbf{p_2}$. This
amounts to take equal diagonal blocks in the right side of
Eq.~(\ref{bipartite-block-structure}), with no constraint on the
off-diagonal blocks. This second choice results in a free energy with a
slightly different form but leads to the same main results as the
choice of Eq.~(\ref{bipartite-block-structure}). This alternative form
is discussed further in the Appendix.

Combining the requirements imposed by the three symmetries, we arrive
at an interaction matrix of the form
\begin{eqnarray}
  H_\mathrm{int} = \sum_{\mu=0}^3 \sigma_\mu \left(\begin{array}{cccc} 
B_{\mu d} & B_{\mu o} & C_{\mu d} & C_{\mu o}\\
B_{\mu o} & B_{\mu d} & C_{\mu o} & C_{\mu d} \\
C^*_{\mu d} & C_{\mu o}^* & B_{\mu d}^* & B_{\mu o}^* \\
C^*_{\mu o} & C_{\mu d}^* & B_{\mu o}^* & B_{\mu d}^*\\
\end{array}\right),
\label{Hint}
\end{eqnarray}
where the $4\times 4$ block structure of $H_{\mathrm{int}}$ refers to
the regions of the Brillouin zone. Each block is described by an $M
\times M$ matrix; blocks $B_{\mu d}$ and $B_{\mu o}$ are Hermitian,
while $C_{\mu d}$ and $C_{\mu o}$ are complex symmetric. Overall,
$H_{\mathrm{int}}$ contains $16$ independent blocks.

{\bf Probability distribution}.  We represent the $16$ blocks by $A_b$
with $b=1,\ldots,16$. Their matrix elements are drawn on a normal
distribution
\begin{eqnarray}
 P(H_{\mathrm{int}}) = \exp\left(- 8 M \sum^{16}_{b=1}
 \Sigma_{b}^2 
\mathrm{Tr}\left(A_b^{\phantom{\dagger}}A_b^{\dagger}\right)\right),
\label{PH-Ab}
\end{eqnarray}
where $\Sigma_{b}^2$ represent inverse variances. This form allows us
to perform the integration over $H_{\mathrm{int}}$ analytically and thus
to determine the partition function exactly.\footnote{As is usually the
  case in random matrix theory, the technically convenient choice of a
  Gaussian distribution is completely passive.  All final results will
  depend only on the inverse variances, $\Sigma_{b}^2$, and identical
  results would be obtained for other choices of $P(H_{\mathrm int})$
  with equal variances.}  In order to make the partition function
invariant under $SU(2)$ rotations, the inverse variances associated
with each of the three blocks that describe a spin fluctuation
exchange are chosen equal. Since 12 of the 16 independent blocks
describe spin-fluctuations and four describe density fluctuations, we
arrive at a total of $4+4 = 8$ independent variances.  The resulting
distribution function is then given as
\begin{eqnarray}
P(H_{\mathrm{int}}) & = & \exp\left( - 8 M \left(
\Sigma^2_{B0d}\mathrm{Tr}(B_{0d}B_{0d}^\dagger) +
\Sigma^2_{\mathbf{B}_d}\mathrm{Tr}(\mathbf{B}_{d}\cdot \mathbf{B}_{d}^\dagger)
\right.\right.\nonumber \\ &&+
\Sigma^2_{B0o}\mathrm{Tr}(B_{0o}B_{0o}^\dagger) +
\Sigma^2_{\mathbf{B}_o}\mathrm{Tr}(\mathbf{B}_{o}\cdot\mathbf{B}_{o}^\dagger) +
\Sigma^2_{C0d}\mathrm{Tr}(C_{0d}C_{0d}^\dagger) +
\Sigma^2_{\mathbf{C}_d}\mathrm{Tr}(\mathbf{C}_{d}\cdot\mathbf{C}_{d}^\dagger)
\nonumber \\ &&+ \left.\left.
\Sigma^2_{C0o}\mathrm{Tr}(C_{0o}C_{0o}^\dagger) +
\Sigma^2_{\mathbf{C}_o}\mathrm{Tr}(\mathbf{C}_{o}\cdot\mathbf{C}_{o}^\dagger)
\right)\right),
\label{PH-final}
\end{eqnarray}
where the inverse variances can be tuned individually at will in order
to favor various scattering mechanisms.

{\bf Non-interacting terms}. The non-interacting part of the single
particle Hamiltonian is written as
\begin{eqnarray}
  \psi^\dagger H_0 \psi = \psi^\dagger \left(- \mu + {\Omega}_T +
  \Gamma_t \right)\psi^{\phantom{\dagger}}.
\label{H0}
\end{eqnarray}
Here, the chemical-potential term, $\mu$, is a scalar while 
${\Omega_T}$ and ${\Gamma_t}$ are matrices which describe temperature
dependence and hopping terms, respectively.

Temperature is introduced via Matsubara frequencies.  Following
previous work in QCD,\cite{VanJac00, VanJac00b} we include only the
two lowest frequencies, $\pm i \pi T$. With this approximation, 
${\Omega}_T$ takes the form
\begin{eqnarray}
{\Omega_T} = \mathrm{diag}(i \pi T, -i \pi T) \otimes
(\mathbf{1})_{\mathrm{spin}}\otimes (\mathbf{1})_Q \otimes
(\sigma_3)_P,
\label{T-term}
\end{eqnarray}
where the first term on the right side is an $M\times M$
diagonal matrix. Hence, in each momentum region, half the states have
a positive frequency while the other half have a negative frequency. The
final term on the right side of Eq.~(\ref{T-term}) serves to
implement the parametrization introduced above: states with a fixed
label $i$ in regions $1$ and $2$ have frequencies opposite to the
corresponding states in regions $3$ and $4$. 

Although limiting the sum over Matsubara frequencies leads to an
oversimplified description of temperature dependence, we believe 
it to be sufficient to determine the general characteristics of the phase 
transition. In fact, we will see below that the parameter $T$ serves 
as an energy scale which influences the energy balance between the 
various order parameters, much as the average thermal energy does 
when all frequencies are taken into account.  The inclusion of all 
frequencies would certainly modify the resulting phase diagram but 
only in the trivial sense that every temperature $T$ is mapped 
monotonically to a new value.\cite{VanJac01}  We will not seek 
to refine the $T$-dependence here since such mapping does not
alter the phase topology, i.e., the occurrence and the order of
transition lines. Since $T$ as introduced here is an arbitrary 
temperature scale, we will also drop the factor $\pi$ to simplify 
notation.

The hopping term $\Gamma_t$ in Eq.~(\ref{H0}) is written as 
\begin{eqnarray}
\Gamma_t = \mathrm{diag}(t, -t)\otimes
(\mathbf{1})_{\mathrm{spin}} \otimes (\sigma_3)_Q \otimes
(\mathbf{1})_P,
\label{Gammat}
\end{eqnarray}
where the first term on the right side is a diagonal $M\times M$ matrix
containing $M/2$ elements of $+t$ and $M/2$ elements of $-t$. The
matrix $\Gamma_t$ mimics the nearest-neighbor hopping energy
\begin{eqnarray}
  \xi_\mathbf{p} = -2 t_0 \left(\cos(p_x a/\hbar)+\cos(p_y a/\hbar)\right),
\label{xi}
\end{eqnarray}
whose band is symmetric around $\xi_\mathbf{0} = 0$ and which
satisfies $\xi_{\mathbf{p+Q}}= - \xi_\mathbf{p}$. Here, again, we
neglect the detailed momentum dependence of the hopping energy and
retain only its symmetries around $\xi_\mathbf{0} = 0$ and under
a shift by $\mathbf{Q}$. In Eq.~(\ref{Gammat}), $t$ is a measure of
the strength of the hopping term, which can be tuned against the
variances and thus against the interaction strength.

\section{Thermodynamic potential}
\label{s:TP}

In this section we evaluate the thermodynamic potential corresponding
to the partition function of Eq.~(\ref{partition-function}) using
methods which are standard in random matrix
theory.\cite{JacksonVerbaarschot96, VanJac00} It is worth emphasing
that the random matrix model of Eq. (12) is solvable, i.e., the
saddle-point method to be discussed below becomes exact in the
thermodynamic limit.  We will concentrate on the main results and
present more detailed calculations in the Appendix. These calculations
are performed in three steps. The first step consists of integrating
over the interaction matrix elements according to the distribution
$P(H_\mathrm{int})$ of Eq.~(\ref{PH-final}). This integration leads to
a four-fermion interaction, $Y$, which receives contributions from the
eight independent blocks of $H_\mathrm{int}$ in Eq.~(\ref{Hint}). In
the second step, the fermion fields are rearranged and the terms of
$Y$ are put in two groups.  The first group contains products of
bilinears of the form $\sim\psi^\dagger_i\psi_i^{\phantom{\dagger}}$
(with an implicit sum over $i$), which includes the antiferromagnetic
order parameter.  The second group contains products involving
bilinears of the form $\psi_i\psi_i$, which are relevant for
superconductivity.  To keep track of momentum and spin indices while
rearranging terms, we use the generalized Fierz identities of the
Appendix.  In the final step, quartic fermion terms are linearized by
means of Hubbard-Stratonovitch transformations, which introduce an
auxiliary real field $\sigma$, to be associated with an
antiferromagnetic order parameter, and a complex field $\Delta$, to be
associated with superconductivity. The resulting form of the partition
function is given as
\begin{eqnarray}
  Z(\mu,T) = \int d\sigma d\Delta d\Delta^* e^{-8 M \Omega(\sigma,\Delta)},
\label{Z-final}
\end{eqnarray}
where $\Omega$ is the thermodynamic potential
\begin{eqnarray}
  \Omega(\sigma,\Delta) & = & A |\Delta|^2 + B\sigma^2 - {1\over 4}\,
\log((\sqrt{\sigma^2+t^2}-\mu)^2+|\Delta|^2+ T^2) \nonumber\\
&& - {1\over 4}\,
\log((\sqrt{\sigma^2+t^2}+\mu)^2+|\Delta|^2+ T^2),
\label{TP-final}
\end{eqnarray}
with 
\begin{eqnarray}
  A & \equiv & {8} \left({1\over \Sigma_{B0d}^2} - {1\over \Sigma_{B0o}^2}
  + {1\over \Sigma_{C0d}^2} - {1\over \Sigma_{C0o}^2} - {3\over
    \Sigma_{\mathbf{B}_d}^2} + {3\over \Sigma_{\mathbf{B}_o}^2} - {3\over
    \Sigma_{\mathbf{C}_d}^2} + {3\over \Sigma_{\mathbf{C}_o}^2}\right)^{-1},
\label{Atext}\\
 B & \equiv & 8 \left(- {1\over \Sigma_{B0d}^2} - {1\over
   \Sigma_{B0o}^2} - {1\over \Sigma_{C0d}^2} - {1\over \Sigma_{C0o}^2}
 + {1\over \Sigma_{\mathbf{B}_d}^2} +  {1\over \Sigma_{\mathbf{B}_o}^2} + {1\over
   \Sigma_{\mathbf{C}_d}^2} + {1\over
   \Sigma_{\mathbf{C}_o}^2}\right)^{-1}.
\label{Btext}
\end{eqnarray}
Equations (\ref{TP-final})-(\ref{Btext})
constitute the main result of the model.  Note that the
factor $8M$ in the argument of the exponential in~Eq.~(\ref{Z-final})
plays the role of the volume of the system since $M$ scales with the
total number of lattice sites. The phase diagram can be established
with the help of the saddle-point approximation to Eq.~(\ref{Z-final}), 
which requires the simultaneous solution of the two gap equations 
$\partial \Omega/\partial\sigma = 0$ and $\partial\Omega/\partial\Delta 
= 0$. Due to the logarithm term in the right side of Eq.~(\ref{TP-final}), 
these equations are polynomial in the auxiliary fields and can be solved 
analytically. In the thermodynamic limit $M\to \infty$, the saddle-point 
approximation becomes exact and the solutions of the gap equations that 
achieve the lowest value for $\Omega$ describe the thermodynamic 
phases of the system.

As mentioned in the introduction, the model becomes increasingly simple 
as one proceeds through the calculation.  From the initially large number 
of parameters required to describe $H_\mathrm{int}$ in Eq.~(\ref{Hint}), 
there remain only a few constants in the final form of the 
thermodynamic potential [Eq.~(\ref{TP-final})]. This potential 
has a remarkably simple
structure.  Anticipating the results of the following section,
we note first that the topology of the phase diagram depends only on 
a single parameter ratio
\begin{eqnarray}
  \alpha = \frac{B}{A},
\label{alpha}
\end{eqnarray}
whose strength characterizes the relative importance of
superconductivity and antiferromagnetism. The terms $A|\Delta|^2$ and
$B\sigma^2$ represent the energy cost of creating a constant field in
the corresponding channel. As shown in the Appendix, the logarithmic
term in $\Omega$ corresponds to the determinant of the Hamiltonian for
a single fermion in fixed constant external fields $\sigma$ and
$\Delta$. These terms have the generic form
\begin{eqnarray}
  \sim \sum_{\varepsilon_\pm}\log\left(i T - \varepsilon_\pm\right)
  \left(-i T - \varepsilon_\pm\right),
\end{eqnarray}
where $\varepsilon_\pm$ are the quasiparticle energies
\begin{eqnarray}
  \varepsilon_\pm = \left(\left(\sqrt{t^2+\sigma^2}\pm\mu\right)^2+
|\Delta|^2\right)^{1/2}.
\label{qp}
\end{eqnarray}
These expressions are strongly reminiscent of the quasiparticle
energies of earlier mean-field models.\cite{Kyung00, Kyung2002}
Neglecting the triplet order and approximating the square of the form
factor as $\phi_d^2 \approx 1$ in Eqs.~(10) and (11) of
Ref.\cite{Kyung00}, one finds quasiparticle energies of the form
$E_\pm(\mathbf{p}) =((
\sqrt{\xi_\mathbf{p}^2+\sigma^2}\pm\mu)^2+|\Delta|^2)^{1/2}$ with
$\sigma=2 J m$ and $\Delta = J d$. Equation~(\ref{qp}) shows a result
of similar structure with, however, the simplification
$\xi_\mathbf{p}\sim t$ which is a consequence of our coarse
description of momentum states which leads us to ignore the detailed
momentum dependence of kinetic-energy terms.

Thus, the basic structure of the potential $\Omega$ is simply related
to the energies of the elementary excitations of the system for fixed
constant external fields. This means that the thermodynamic potential
could have been constructed immediately from the 
knowledge of the quasiparticle energies alone, without going through
the steps described in Sec.~\ref{s:RM}.  Note, however, that the
additional information which comes from constructing the interactions
at the more microscopic level is useful.  Through the dependence of
$A$ and $B$ on the individual variances, it establishes connections
between the microscopic mechanisms and the global properties of the
system.

\section{Phase diagram}
\label{s:PD}

We now consider the various topologies that can be realized in the
phase diagram. Despite the simplicity of the thermodynamic potential, a
 full exploration of the parameter space is a
considerable task. Thus, we concentrate on a restricted number of
physically relevant cases.

First, we assume that the interactions are attractive in the
antiferromagnetic channel. As shown in the Appendix,
this requires that the variances satisfy the inequality
\begin{eqnarray}
   B & = & 8 \left(- {1\over \Sigma_{B0d}^2} - {1\over \Sigma_{B0o}^2} -
{1\over \Sigma_{C0d}^2} - {1\over \Sigma_{C0o}^2}
+ {1\over \Sigma_{\mathbf{B}_d}^2} + {1\over \Sigma_{\mathbf{B}_o}^2}+ 
{1\over \Sigma_{\mathbf{C}_d}^2} + {1\over \Sigma_{\mathbf{C}_o}^2}\right)^{-1} > 0,
\label{ineq-AF}
\end{eqnarray}
which implies that spin-fluctuation exchange is stronger than
density-fluctuation exchange. In fact, this condition can be related
to the requirement that the on-site potential in the Hubbard model is
repulsive. In momentum representation, the Hubbard Hamiltonian is
written as
\begin{eqnarray}
  H_{\mathrm{H}} = \sum_{\alpha\mathbf{p}} \varepsilon_\mathbf{p}
  \psi^\dagger_{\alpha\mathbf{p}}
  \psi^{\phantom{\dagger}}_{\alpha\mathbf{p}} + U
  \sum_{\mathbf{p}_1\mathbf{p}_2\mathbf{p}_3\mathbf{p}_4}
  \delta_{\mathbf{p}_1+\mathbf{p}_3,\mathbf{p}_2+\mathbf{p}_4} \,
  \psi^\dagger_{\uparrow\mathbf{p}_1}
  \psi^{\phantom{\dagger}}_{\uparrow\mathbf{p}_2}
  \psi^\dagger_{\downarrow\mathbf{p}_3}
  \psi^{\phantom{\dagger}}_{\downarrow\mathbf{p}_4},
\end{eqnarray}
with $U$ positive. The corresponding partition function contains the
thermal average $\mathrm{Tr}(e^{-H_\mathrm{H}/T}\ldots)$, where the
equivalent of the weighting factor $e^{-H_\mathrm{H}/T}$ in the random
matrix model is the term $e^Y$. Hence, a \emph{positive} $U$ in the Hubbard
model would correspond to a \emph{negative} four-fermion term of the
form $\psi^\dagger_{\uparrow i}\psi^{\phantom{\dagger}}_{\uparrow
  j}\psi^\dagger_{\downarrow j} \psi^{\phantom{\dagger}}_{\downarrow
  i}$ in the random matrix model.  Close inspection of the
four-fermion potentials in Eqs.~(\ref{YB0d})--(\ref{YCveco}) reveals that
they contain such terms and that they will be negative provided
the inequality of Eq.~(\ref{ineq-AF}) is satisfied.

As a second assumption, we consider interactions which are attractive
in the superconducting $d$-wave channel but repulsive in the $s$-wave
channel. This requires a particular choice of the inverse variances.
In the Appendix, we show that such an interaction can be found among
those that favor spin-fluctuations by putting more statistical weight
on the blocks $\mathbf{B}_o$ and $\mathbf{C}_o$ than on $\mathbf{B}_d$
and $\mathbf{C}_d$. This choice leads to a positive constant $A$ in
Eq.~(\ref{Atext}). Such a choice of the variances is expected to
result from interactions which favor large momentum transfer of order
$\sim \mathbf{Q}$. This result can be related to the antiferromagnetic
spin-fluctuation model of Ref.\cite{Chubukov2002} in which the spin
susceptibility is peaked at a momentum exchange $\sim \mathbf{Q}$. In
the present approach, however, we do not attempt to describe the
dynamics of spin exchange in detail but rather use the coarse device
of the inverse variances to tune the relative strengths of the
exchange mechanisms.

Given the restricted parameter space that results from these two 
assumptions, the system can develop different phase structures in the
($\mu$,$T$) plane as a function of the parameter ratio $\alpha =
B/A$. The phase structures can be grouped according to their
topology. We identify four distinct topologies which emerge as 
the parameter $\alpha$ is gradually increased. The system
switches from one topology to the next at specific values of $\alpha$
that depend on the strength of the hopping term $t$. In general, 
systems with larger $t$ develop a larger kinetic energy per charge 
carrier and are found to favor superconductivity over antiferromagnetism.

It is useful to note that because our model only includes
nearest-neighbor hopping, it cannot distinguish between hole- and
electron-doped systems. This symmetry can be seen in the potential of
Eq.~(\ref{TP-final}), where $\Omega$ is an even function of
$\mu$. Hence, $\mu = 0$ corresponds here to half-filling. In
principle, next-nearest neighbor and higher-order hopping terms 
could be added to the model by using a hopping matrix, $\Gamma_t$, whose
eigenvalues have a sign distribution that reproduces the symmetries of
the corresponding kinetic energies in the various regions of the
Brillouin zone.  Although we have limited ourselves to nearest-neighbor 
terms here, we expect that a model with a more elaborate hopping matrix 
would distinguish between electron and hole dopings.

\subsection{Antiferromagnetism alone}
\label{ss:AFalone}

For the smallest values of $\alpha=B/A$, superconductivity is too weak
to compete with antiferromagnetism. Such a situation corresponds, for
instance, to a large value for $A$, leading to a prohibitively large
energy cost $\sim A |\Delta|^2$ for creating a constant field
$\Delta$.

Consider first the system at zero temperature and at half-filling
($\mu = 0$).  Setting $\Delta = 0$ in Eq.~(\ref{TP-final}), we find
the gap equation 
\begin{eqnarray}
\left. {\partial \Omega \over \partial \sigma} \right|_{\Delta=0,\mu=0,T=0} = 
2 B \sigma - \frac{\sigma}{\sigma^2+t^2} = 0,
\end{eqnarray}
which gives either $\sigma = 0$ (paramagnetic phase, PA), or
$\sigma=\sqrt{1/(2 B) - t^2}$ (antiferromagnetic phase, AF). The
latter solution is real if the hopping term is not too strong, $2 B
t^2 < 1$. Given that $B$ scales as an inverse variance [see
Eq.~(\ref{Btext})], this condition is equivalent to the inequality
$t/t_\mathrm{TH} < 1$, where the threshold value
$t_\mathrm{TH}=1/\sqrt{2B}$ is a measure of the interaction
strength. When this condition is fulfilled, $\sigma=\sqrt{1/(2 B) -
  t^2}$ is the absolute minimum of $\Omega$ and the ground state is
antiferromagnetic. In the rest of this work, we will explicitly assume
that $t < t_\mathrm{TH}$, so that the half-filled state is
antiferromagnetic.

In contrast to our approach, mean-field studies of the Hubbard model
find an antiferromagnetic ground state no matter how weak the
interactions are or how large the hopping term is.\cite{Hirsch1985,Lin1987} 
There, the absence of a threshold results from the logarithmic singularity 
that occurs in the density of states at the edge of the magnetic Brillouin 
zone. At half-filling, the divergent density of states leads to a gap with 
an exponential dependence on $t/U$, $\sigma \sim t \exp(-2 \pi \sqrt{t/U})$, 
where $U$ is the strength of the on-site repulsion.  In the random
matrix approach, the detailed band structure is ignored and, because of
the coarse description of momentum states, no divergence appears 
in the density of states. As a result, the interaction must be sufficiently 
strong to produce an antiferromagnetic ground state. Given the level of 
approximations underlying our approach, this behavior is not unreasonable.  
Discrepancies with microscopic theories are to be expected in cases where 
the condensates are weak and thus sensitive to fluctuations. Understanding the 
fate of these condensates clearly requires more than mean-field approximation.

\begin{figure}[ht]

\begin{picture}(0,0)%
\includegraphics{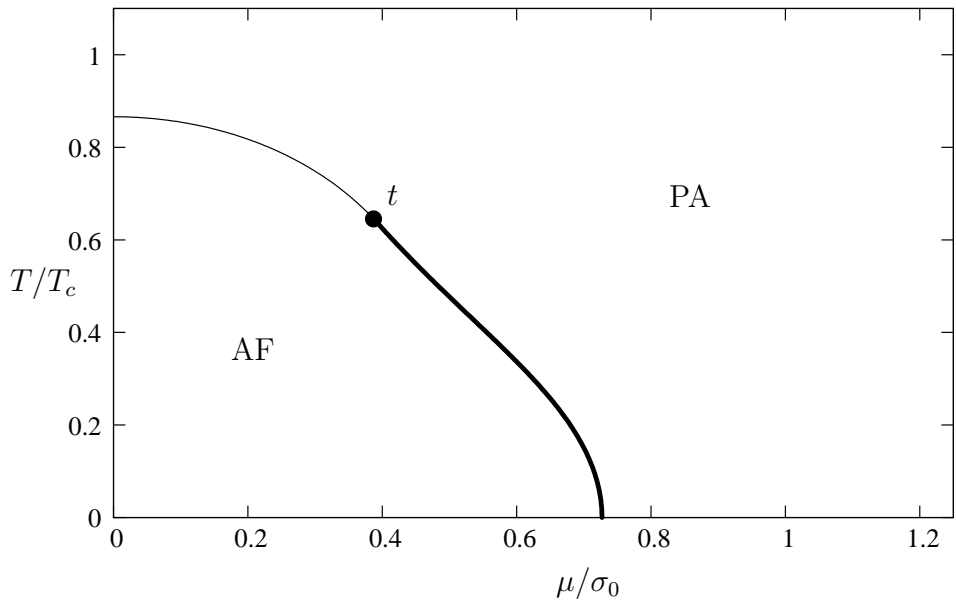}
\end{picture}%
\setlength{\unitlength}{3947sp}%
\begingroup\makeatletter\ifx\SetFigFontNFSS\undefined%
\gdef\SetFigFontNFSS#1#2#3#4#5{%
  \reset@font\fontsize{#1}{#2pt}%
  \fontfamily{#3}\fontseries{#4}\fontshape{#5}%
  \selectfont}%
\fi\endgroup%
\begin{picture}(5952,3753)(1048,-4252)
\put(2451,-2719){\makebox(0,0)[lb]{\smash{{\SetFigFontNFSS{12}{14.4}{\familydefault}{\mddefault}{\updefault}{\color[rgb]{0,0,0}AF}%
}}}}
\put(5203,-1759){\makebox(0,0)[lb]{\smash{{\SetFigFontNFSS{12}{14.4}{\familydefault}{\mddefault}{\updefault}{\color[rgb]{0,0,0}PA}%
}}}}
\put(4493,-4174){\makebox(0,0)[lb]{\smash{{\SetFigFontNFSS{12}{14.4}{\familydefault}{\mddefault}{\updefault}{\color[rgb]{0,0,0}$\mu/\sigma_0$}%
}}}}
\put(1063,-2289){\makebox(0,0)[lb]{\smash{{\SetFigFontNFSS{12}{14.4}{\familydefault}{\mddefault}{\updefault}{\color[rgb]{0,0,0}$T/T_c$}%
}}}}
\put(3432,-1748){\makebox(0,0)[lb]{\smash{{\SetFigFontNFSS{12}{14.4}{\familydefault}{\mddefault}{\updefault}{\color[rgb]{0,0,0}$t$}%
}}}}
\end{picture}%

  \caption{Phase diagram for $t/t_\mathrm{TH}=0.5$ and $\alpha <
    0.1$. The transition from the antiferromagnetic phase (AF) to the
    paramagnetic phase (PA) is second-order at half-filling (thin
    line) and first-order at zero temperature (thick line). These two
    lines merge at a tricritical point, $t$. Here, temperature is
    plotted in units of $T_c \equiv T_c(\mu=0,t=0)$, which is the
    transition temperature at half-filling in the limit $t\to 0$,
    while chemical potential is plotted in units of
    $\sigma_0\equiv1/\sqrt{2B}$, which represents the AF field at
    half-filling, zero-temperature, and for $t\to 0$.}
  \label{f:AFalone-t0dot5}
\end{figure}

We now turn to the phase diagram in the $(\mu,T)$ plane. Choosing
$t/t_{\mathrm{TH}}=0.5$, the superconducting phase does not 
develop so long as $\alpha < \alpha_\Delta$ with $\alpha_\Delta\approx
0.1$. The corresponding phase structure is shown in
Fig.~\ref{f:AFalone-t0dot5} and resembles that of chiral symmetry breaking
in QCD with two flavors and three colors, in the limit where a
color-superconducting phase is ignored (see Ref.\cite{VanJac00}). The
gap equation $\left. d\Omega/d\sigma = 0\right|_{\Delta = 0}$ has a
form similar to the QCD problem. It has a solution with $\sigma = 0$,
describing a paramagnetic phase, and solutions which satisfy the
quadratic equation
\begin{eqnarray}
x^2 + 2 x (-\mu^2  +T^2 - {1\over 4 B}) + (\mu^2 + T^2)^2 +{\mu^2 - T^2
  \over 2 B} = 0,
\label{x}
\end{eqnarray}
where $x=\sigma^2+t^2$. For moderate $\mu$ and high temperature, the
only real solution is $\sigma = 0$ and the system is in a paramagnetic
phase. Decreasing $T$ at fixed $\mu$, two additional real solutions
\begin{eqnarray}
  \sigma = \pm \sigma_{AF} = \pm \left(\mu^2 - t^2 - T^2 +{1\over 4 B}
    + {1\over 4 B} \sqrt{1-64 B^2 \mu^2 T^2}\right)^{1/2}
\end{eqnarray}
can be found below the critical temperature
\begin{eqnarray}
  T_c(\mu,t) = \left({1\over 4 B} - \mu^2 - t^2 + {1 \over 4 B}
  \sqrt{1 - 16 B \mu^2 + 64 B^2 \mu^2 t^2} \right)^{1/2}.
\label{Tc}
\end{eqnarray}
Below $T_c(\mu,t)$, these solutions are local minima of the free
energy $\Omega(\sigma,\Delta=0)$ and $\sigma = 0$ becomes a
maximum. The finite roots $\pm \sigma_{AF}$ describe an
antiferromagnetic phase.  They vanish at $T_c(\mu,t)$, which thus
characterizes a second-order transition from an antiferromagnetic to a
paramagnetic phase. In the opposite regime of low temperatures and
finite $\mu$, the transition is found to be discontinuous. It takes
place along a first-order line [actually, a triple line (see
Ref.\cite{VanJac00})] which starts on the zero-temperature axis at
\begin{eqnarray}
  \mu_1 = \sqrt{{0.14\over B} + t^2},
\label{mu1}
\end{eqnarray}
and extends with decreasing $\mu$ toward the zero-$\mu$ axis. This
line meets the second-order line $T=T_c(\mu,t)$ at the tricricital
point $t \equiv (\mu_3,T_3)$, given as
\begin{eqnarray}
 \mu_3 & = & \left(- \frac{1- 4 B t^2}{8 B} + \frac{\sqrt{1+(1-4 B t^2)^2}}{8 B}
\right)^{1/2},\label{mu3}\\
T_3 & = & \left(\frac{1- 4 B t^2}{8 B} + \frac{\sqrt{1+(1-4 B t^2)^2}}{8 B}
\right)^{1/2}.
\label{T3}
\end{eqnarray}
The first- and second-order lines meet with equal slopes $dT/d\mu$.

The thermodynamic potential $\Omega(\sigma, \Delta=0)$ can be expanded
as a series of powers in $\sigma$ near the critical lines. The result
resembles a Ginzburg-Landau expansion.  For $\mu < \mu_3$ and near
$T_c(\mu,t)$, the free energy is found to scale as $\Omega(\sigma,
\Delta=0) \approx \Omega(\sigma = 0, \Delta = 0) + b_4(\mu,T)
\,\sigma^4 +{\cal O}(\sigma^6)$, so that the second-order phase
transition has the critical exponents of a mean-field $\phi^4$
theory. Near the tricritical point, one has $\Omega(\sigma, \Delta=0)
\approx \Omega(\sigma = 0, \Delta = 0) + b_6(\mu,T)\,\sigma^6 +{\cal
  O}(\sigma^8)$ and the critical exponents are now those of a
mean-field $\phi^6$ theory. Note, however, that the coefficients $b_4$
and $b_6$ here are \emph{known} functions of $\mu$ and $T$.

\subsection{Competition between antiferromagnetism and superconductivity}
\label{ss:SCemergesout}

\begin{figure}[ht]
\begin{center}


\begin{picture}(0,0)%
\includegraphics{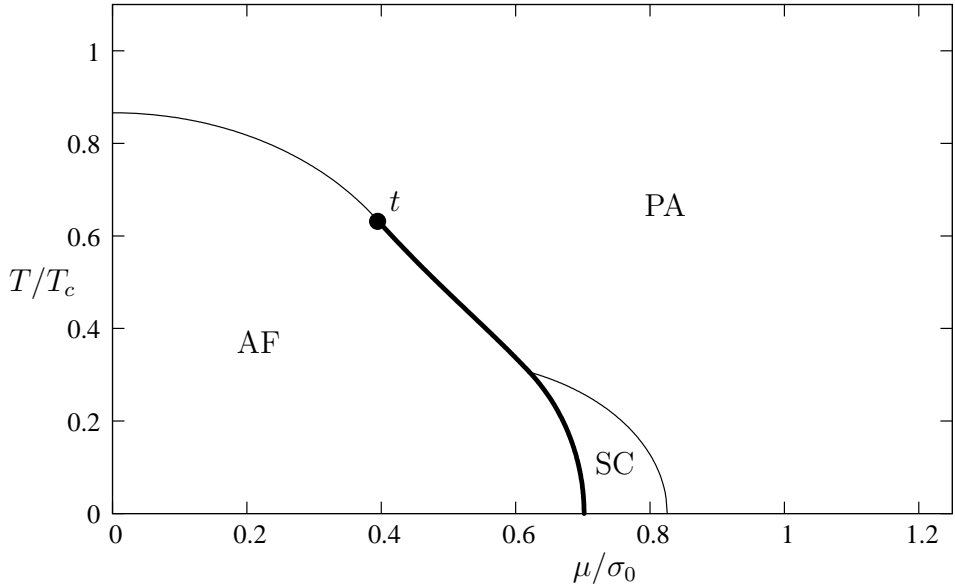}
\end{picture}%
\setlength{\unitlength}{3947sp}%
\begingroup\makeatletter\ifx\SetFigFontNFSS\undefined%
\gdef\SetFigFontNFSS#1#2#3#4#5{%
  \reset@font\fontsize{#1}{#2pt}%
  \fontfamily{#3}\fontseries{#4}\fontshape{#5}%
  \selectfont}%
\fi\endgroup%
\begin{picture}(5952,3683)(1048,-4182)
\put(2493,-2699){\makebox(0,0)[lb]{\smash{{\SetFigFontNFSS{12}{14.4}{\familydefault}{\mddefault}{\updefault}{\color[rgb]{0,0,0}AF}%
}}}}
\put(5053,-1839){\makebox(0,0)[lb]{\smash{{\SetFigFontNFSS{12}{14.4}{\familydefault}{\mddefault}{\updefault}{\color[rgb]{0,0,0}PA}%
}}}}
\put(4743,-3459){\makebox(0,0)[lb]{\smash{{\SetFigFontNFSS{12}{14.4}{\familydefault}{\mddefault}{\updefault}{\color[rgb]{0,0,0}SC}%
}}}}
\put(1063,-2309){\makebox(0,0)[lb]{\smash{{\SetFigFontNFSS{12}{14.4}{\familydefault}{\mddefault}{\updefault}{\color[rgb]{0,0,0}$T/T_c$}%
}}}}
\put(4603,-4104){\makebox(0,0)[lb]{\smash{{\SetFigFontNFSS{12}{14.4}{\familydefault}{\mddefault}{\updefault}{\color[rgb]{0,0,0}$\mu/\sigma_0$}%
}}}}
\put(3454,-1808){\makebox(0,0)[lb]{\smash{{\SetFigFontNFSS{12}{14.4}{\familydefault}{\mddefault}{\updefault}{\color[rgb]{0,0,0}$t$}%
}}}}
\end{picture}%

  \caption{Phase diagram for $t/t_\mathrm{TH}=0.5$ and $\alpha =
    0.2$. In addition to the phase structure of
    Fig.~\ref{f:AFalone-t0dot5}, there is a superconducting phase (SC)
    emerging out of the antiferromagnetic phase via a first-order
    transition. This phase undergoes a second-order transition to the
    paramagnetic phase at either higher $\mu$ or higher $T$. Here, $t$
    is a tricritical point. The scales $T_c$ and $\sigma_0$ are those
    defined in the caption of Fig.~\ref{f:AFalone-t0dot5}.}
  \label{f:SCemergesout-t0dot5-al0dot2}
\end{center}
\end{figure}

As the coupling ratio $B/A$ increases, a superconducting phase can be
favored over an antiferromagnetic one. As doping is increased, the new
carriers disrupt the antiferromagnetic correlations. If the
interaction is sufficiently strong in the pairing channel, this can 
lead to a transition to a superconducting phase with $\Delta \neq
0$. Such a transition can generally take place in one of two ways, 
either through the appearance of a ``wedge'' of mixed broken symmetry 
with both $\sigma\neq 0$ and $\Delta \neq 0$, with continuous transitions
toward the pure AF and SC phases, or via a discontinuous transition
between the two pure phases. The first case was encountered in a random matrix
model of Ref.\cite{VanJac00} when the coupling ratio of QCD was
altered in favor of color superconductivity. Here, however, the second case
is found as shown in Fig.~\ref{f:SCemergesout-t0dot5-al0dot2}.

To understand the onset of superconductivity, consider a pure phase
with $\sigma = 0$ and $\Delta \neq 0$. The gap equation,
$d\Omega(\sigma = 0,\Delta)/d\Delta=0$ always has a root $\Delta = 0$,
which is a local minimum at either large $\mu$ or large $T$. For 
moderate $\mu$ and for $T$ less than
\begin{eqnarray}
  T_{c\Delta}(\mu,t) = \left({1\over 4 A}- \mu^2 - t^2 + {1\over 4 A}
\sqrt{1 + 64 A^2 \mu^2 t^2}\right)^{1/2},
\label{TcDelta}
\end{eqnarray}
the solution $\Delta = 0$ becomes a local maximum, while
$\Omega(0,\Delta)$ exhibits two local minima, given by the
roots with
\begin{eqnarray}
  |\Delta| = \Delta_{SC} = \left({1\over 4 A} - \mu^2 - T^2 -t^2 
+ {1\over 4 A} \sqrt{1+64 A^2\mu^2 t^2}\right)^{1/2},
\end{eqnarray}
which describe the superconducting phase. $\Delta_{SC}$ vanishes on the
curve $T = T_{c\Delta}(\mu,t)$, which is thus a second-order
transition line. In particular, the thermodynamic potential behaves
like $\Omega(\sigma=0,\Delta) \approx \Omega(0,0) + a_4 |\Delta|^4+{\cal
  O}(|\Delta|^6)$ in the vicinity of the phase boundary.

The curve $T=T_{c\Delta}(\mu,t)$ meets the $T=0$ axis at $\mu_{+}
= (1/(4 A)+ t^2 + \sqrt{1+16 A t^2}/(4 A))^{1/2}$.  For $t^2 \le 1/(2
A)$, this line also meets the $\mu=0$ axis, so that the region $T \le
T_{c\Delta}(\mu,t)$ contains the half-filled state with $\mu=T=0$. For
$t^2>1/(2 A)$, on the other hand, the curve $T = T_{c\Delta}(\mu,t)$
has the shape of a dome, which starts on the $T=0$ axis at
$\mu_{-} = (t^2+1/(4 A) - \sqrt{1+16 A t^2}/(4 A))^{1/2}$,
reaches a maximum $T_\mathrm{max} = (1/(8 A t))\sqrt{1 + 16 A t^2}$ at
$\mu_\mathrm{max}= \sqrt{64 A^2 t^4 - 1}/(8 A t)$ and decreases again
to meet the $T=0$ axis at $\mu=\mu_+$.

The condition for the superconducting phase to emerge out of the AF
phase is now clear.  The curve $T=T_{c\Delta}$ must end at $\mu_+ >
\mu_1$, where $\mu_1$ is given by Eq.~(\ref{mu1}). In that case, there
is an intermediate region where the superconducting state achieves a
lower energy than both the paramagnetic and the antiferromagnetic
states. For a fixed ratio $t/t_{\mathrm{TH}}$, the threshold condition
$\mu_+=\mu_1$ gives the critical value $\alpha=\alpha_\Delta$ which
marks the onset of superconductivity. The result is a decreasing
function of $t$, with $\alpha_\Delta \approx 0.28$ for $t = 0$,
$\alpha_\Delta \approx 0.1$ for $t=0.5\,t_\mathrm{TH}$, and $\alpha_\Delta
\approx 0.03$ in the limit $t \to t_\mathrm{TH}$. For values of
$\alpha > \alpha_\Delta$, the superconducting phase develops in a
wedge adjacent to the antiferromagnetic phase, with a first-order
transition toward the antiferromagnetic state at the lower $\mu$ and
a second-order transition toward the paramagnetic phase at the higher
$\mu$, as illustrated in Fig.~\ref{f:SCemergesout-t0dot5-al0dot2}.

\begin{figure}[ht]
\begin{center}


\begin{picture}(0,0)%
\includegraphics{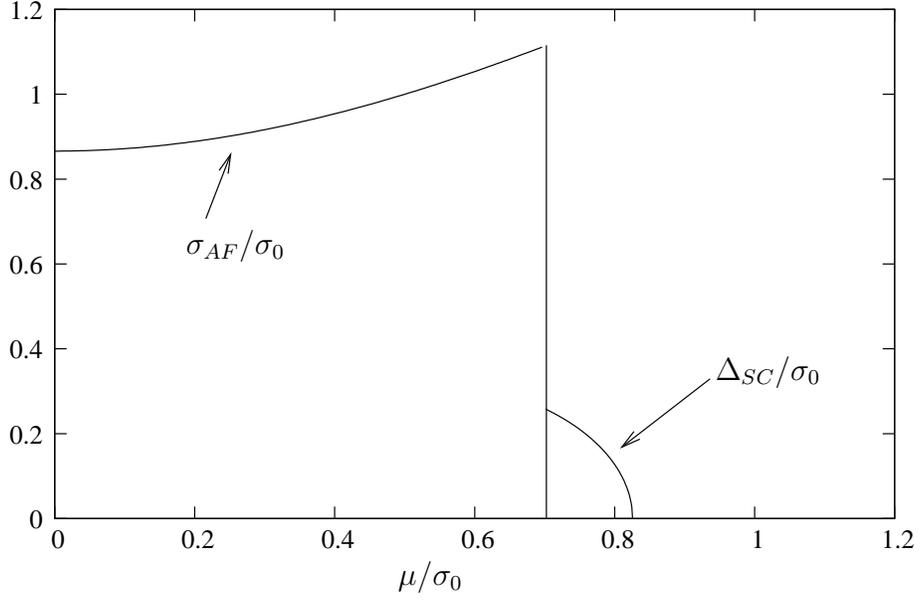}
\end{picture}%
\setlength{\unitlength}{3947sp}%
\begingroup\makeatletter\ifx\SetFigFontNFSS\undefined%
\gdef\SetFigFontNFSS#1#2#3#4#5{%
  \reset@font\fontsize{#1}{#2pt}%
  \fontfamily{#3}\fontseries{#4}\fontshape{#5}%
  \selectfont}%
\fi\endgroup%
\begin{picture}(5726,3752)(1392,-4205)
\put(2539,-2049){\makebox(0,0)[lb]{\smash{{\SetFigFontNFSS{12}{14.4}{\familydefault}{\mddefault}{\updefault}{\color[rgb]{0,0,0}$\sigma_{AF}/\sigma_0$}%
}}}}
\put(5863,-2840){\makebox(0,0)[lb]{\smash{{\SetFigFontNFSS{12}{14.4}{\familydefault}{\mddefault}{\updefault}{\color[rgb]{0,0,0}$\Delta_{SC}/\sigma_0$}%
}}}}
\put(3870,-4127){\makebox(0,0)[lb]{\smash{{\SetFigFontNFSS{12}{14.4}{\familydefault}{\mddefault}{\updefault}{\color[rgb]{0,0,0}$\mu/\sigma_0$}%
}}}}
\end{picture}%

  \caption{Zero-temperature auxiliary fields as a function of $\mu$
    for the parameters corresponding to the phase diagram of
    Fig.~\ref{f:SCemergesout-t0dot5-al0dot2}. The scale 
    $\sigma_0$ is defined in the caption of
    Fig.~\ref{f:AFalone-t0dot5}.}
  \label{f:fields}
\end{center}
\end{figure}

Figure~\ref{f:fields} shows the zero-temperature auxiliary fields
$\sigma_{AF}$ and $\Delta_{SC}$ as a function of the chemical
potential, $\mu$. Note that $\Delta_{SC}$ vanishes at $\mu=\mu_+$ with
the mean-field exponent $1/2$ so that $\Delta_{SC} \sim (\mu_+
-\mu)^{1/2}$. The variation of the antiferromagnetic field
$\sigma_{AF}$ with $\mu$ should be not be considered significant; it is
a direct consequence of the approximate description of temperature
dependence. In fact, taking the sum over all Matsubara frequencies
produces a constant condensation field by a mechanism similar to that
observed in the phase diagram of QCD with two colors and light masses
(see Ref.\cite{VanJac01}).

\subsection{Bicritical point}
\label{ss:bicritical}

\begin{figure}[ht]
\begin{center}

\begin{picture}(0,0)%
\includegraphics{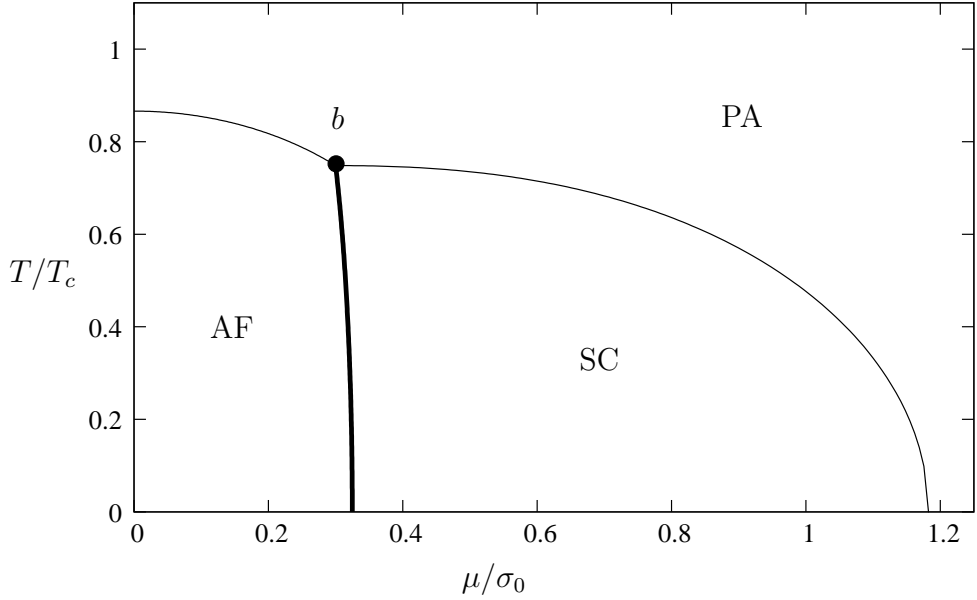}
\end{picture}%
\setlength{\unitlength}{3947sp}%
\begingroup\makeatletter\ifx\SetFigFontNFSS\undefined%
\gdef\SetFigFontNFSS#1#2#3#4#5{%
  \reset@font\fontsize{#1}{#2pt}%
  \fontfamily{#3}\fontseries{#4}\fontshape{#5}%
  \selectfont}%
\fi\endgroup%
\begin{picture}(6082,3763)(918,-4262)
\put(5403,-1289){\makebox(0,0)[lb]{\smash{{\SetFigFontNFSS{12}{14.4}{\familydefault}{\mddefault}{\updefault}{\color[rgb]{0,0,0}PA}%
}}}}
\put(2193,-2619){\makebox(0,0)[lb]{\smash{{\SetFigFontNFSS{12}{14.4}{\familydefault}{\mddefault}{\updefault}{\color[rgb]{0,0,0}AF}%
}}}}
\put(4513,-2819){\makebox(0,0)[lb]{\smash{{\SetFigFontNFSS{12}{14.4}{\familydefault}{\mddefault}{\updefault}{\color[rgb]{0,0,0}SC}%
}}}}
\put(933,-2269){\makebox(0,0)[lb]{\smash{{\SetFigFontNFSS{12}{14.4}{\familydefault}{\mddefault}{\updefault}{\color[rgb]{0,0,0}$T/T_c$}%
}}}}
\put(3773,-4184){\makebox(0,0)[lb]{\smash{{\SetFigFontNFSS{12}{14.4}{\familydefault}{\mddefault}{\updefault}{\color[rgb]{0,0,0}$\mu/\sigma_0$}%
}}}}
\put(2952,-1306){\makebox(0,0)[lb]{\smash{{\SetFigFontNFSS{12}{14.4}{\familydefault}{\mddefault}{\updefault}{\color[rgb]{0,0,0}$b$}%
}}}}
\end{picture}%

  \caption{Phase diagram for $t/t_\mathrm{TH}=0.5$ and $\alpha =
    0.8$. The superconducting phase has grown beyond the tricritical
    point of Fig.~\ref{f:SCemergesout-t0dot5-al0dot2}. The two
    second-order lines (thin lines) separating, respectively, the
    antiferromagnetic and superconducting phases from the paramagnetic
    phase now meet at a bicritical point, $b$, which is also the end
    point of a first-order line (thick line) between the AF and SC
    phases. The scales $T_c$ and $\sigma_0$ are those defined in the
    caption of Fig.~\ref{f:AFalone-t0dot5}.}
  \label{f:bi-t0dot5-al0dot8}
\end{center}
\end{figure}

As $\alpha=B/A$ increases above $\alpha_\Delta$, the superconducting
phase boundary slides up along the first-order transition line between
the antiferromagnetic and the paramagnetic phases. At a new critical
value, $\alpha=\alpha_b$, the superconducting phase boundary reaches
the tricritical point. The value of $\alpha_b$ is readily determined
from the condition $T_{c\Delta}(\mu_3, t) = T_3$ where $\mu_3$, $T_3$,
and $T_{c\Delta}$ are, respectively, given by Eqs.~(\ref{mu3}),
(\ref{T3}), and (\ref{TcDelta}). For $t\to 0$, this gives $\alpha_b =
\sqrt{2}/2$; $\alpha_b$ then decreases with $t$, is equal to $\alpha_b
\approx 0.62$ for $t = 0.5\,t_\mathrm{TH}$ and reaches $\alpha_b \approx
0.29$ in the limit $t \to t_{\mathrm{TH}}$.

For $\alpha > \alpha_b$, the topology of the phase structure is changed.  
The two second-order lines $T_c(\mu,t)$, Eq.~(\ref{Tc}), and $T_{c\Delta}(\mu,
t)$, Eq.~(\ref{TcDelta}), now intersect at a new critical point, $b$, with
\begin{eqnarray}
  \mu_b & = & \left({\alpha \over 4 B} - {\alpha^2 \over 4 B} + t^2 (1
  - \alpha)^2 \right)^{1/2},\label{mub}\\
  T_b & = & \left({\alpha \over 4 B} + {\alpha^2 \over 4 B} - 
\alpha^2 t^2 \right)^{1/2}.\label{Tb}
\end{eqnarray}
In the vicinity of $b$, the thermodynamic potential can be expanded
as $\Omega(\sigma, \Delta) \approx \Omega(0,0) + a_4 \, |\Delta|^4 +
b_4 \, \sigma^4 + c_4 \,\sigma^2 |\Delta|^2$, where $a_4$, $b_4$, and
$c_4$ are known coefficients satisfying the inequality $4\,a_4 b_4 -
c_4^2 < 0$. As a result, the global minimum of $\Omega$ can only be
realized by a pure phase --- either paramagnetic, antiferromagnetic,
or superconducting.  A mixed-broken symmetry state with both
non-vanishing $\sigma$ and $\Delta$ cannot be a global
minimum. Therefore, $(\mu_b,T_b)$ is a bicritical point which ends a
first-order line separating the antiferromagnetic and 
superconducting phases. The coupled gap equations actually have a root
with a mixed-broken symmetry. We have verified that, away from $b$ and
at lower temperatures, (i) this state is always metastable (i.e., it
is a saddle point) whenever its fields are real and (ii) the
transition line between the antiferromagnetic and the superconducting
phases is first-order all the way to the $T=0$ axis.  The resulting
phase structure is shown in Fig.~\ref{f:bi-t0dot5-al0dot8} for the
case $t=0.5~t_\mathrm{TH}$ and $\alpha = 0.8$.

For even larger values of $\alpha$, the bicritical point migrates
toward the $\mu = 0$ axis and reaches it when $\alpha = 1$. In this
case, $T_b = T_c(\mu=0,t)$. The antiferromagnetic phase thus exists 
only at half-filling, and the superconducting phase develops for 
all finite $\mu$ in the region $T<T_{c\Delta}(\mu, t)$.  For $\alpha > 1$,
the superconducting phase wins over the antiferromagnetic state, which
disappears from the phase diagram.

\subsection{Discussion}
\label{ss:discussion}

{\bf Symmetries of $H_\mathrm{int}$ and correlations}.  The
interaction part of the Hamiltonian was constructed by imposing three
different symmetries, which provide the correlations necessary 
for generating finite condensates. Spin rotational and time-reversal
symmetries are intimately related to magnetism.\cite{Flint2008}  The
spin symmetry is obviously necessary in order to express the order
parameters. Time-reversal symmetry leads to a block structure which
produces terms of the form $\psi^\dagger_i\psi_j\psi^\dagger_i\psi_j$,
thereby yielding pairing condensates $\sim \psi_j\psi_j$ and
$\psi^\dagger_i\psi^\dagger_i$. The bipartite block-structure induces
correlations among states whose momenta are separated by
${\mathbf{Q}}$. The resulting four-fermion potential, $\sim
\psi^\dagger_i\psi_j\psi^\dagger_j\psi_i$, thus yields condensates of
the form $\sim \psi^\dagger_i\psi_i$ where the coupled momenta are
separated by $\mathbf{Q}$.

\begin{figure}[ht]
\begin{center}

\begin{picture}(0,0)%
\includegraphics{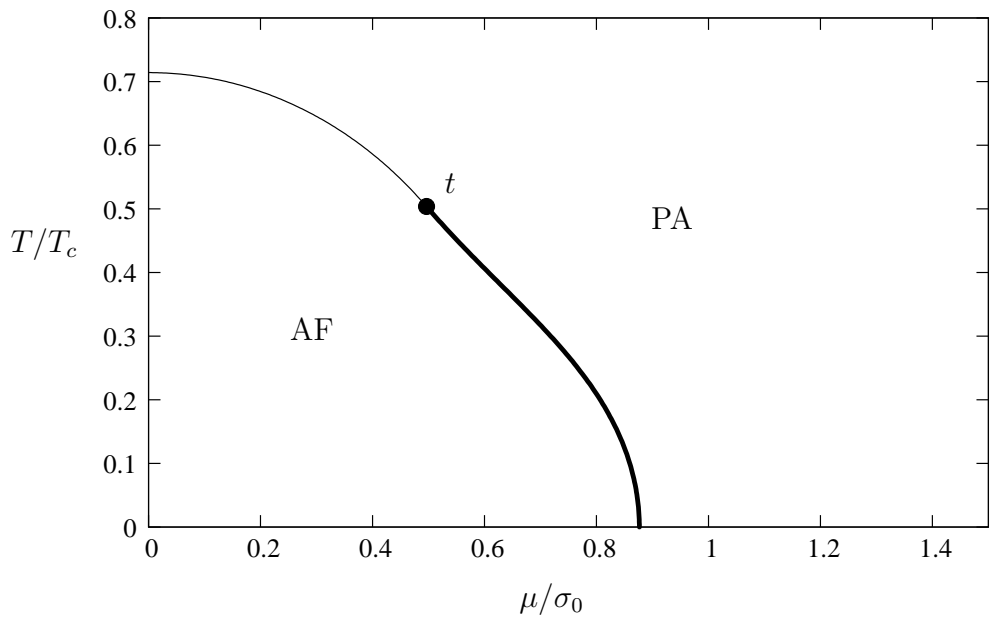}
\end{picture}%
\setlength{\unitlength}{3947sp}%
\begingroup\makeatletter\ifx\SetFigFontNFSS\undefined%
\gdef\SetFigFontNFSS#1#2#3#4#5{%
  \reset@font\fontsize{#1}{#2pt}%
  \fontfamily{#3}\fontseries{#4}\fontshape{#5}%
  \selectfont}%
\fi\endgroup%
\begin{picture}(6168,3839)(832,-4292)
\put(847,-1981){\makebox(0,0)[lb]{\smash{{\SetFigFontNFSS{12}{14.4}{\familydefault}{\mddefault}{\updefault}{\color[rgb]{0,0,0}$T/T_c$}%
}}}}
\put(4042,-4214){\makebox(0,0)[lb]{\smash{{\SetFigFontNFSS{12}{14.4}{\familydefault}{\mddefault}{\updefault}{\color[rgb]{0,0,0}$\mu/\sigma_0$}%
}}}}
\put(3574,-1611){\makebox(0,0)[lb]{\smash{{\SetFigFontNFSS{12}{14.4}{\familydefault}{\mddefault}{\updefault}{\color[rgb]{0,0,0}$t$}%
}}}}
\put(2601,-2536){\makebox(0,0)[lb]{\smash{{\SetFigFontNFSS{12}{14.4}{\familydefault}{\mddefault}{\updefault}{\color[rgb]{0,0,0}AF}%
}}}}
\put(4870,-1834){\makebox(0,0)[lb]{\smash{{\SetFigFontNFSS{12}{14.4}{\familydefault}{\mddefault}{\updefault}{\color[rgb]{0,0,0}PA}%
}}}}
\end{picture}%

  \caption{Phase diagram for $t/t_\mathrm{TH}=0.7$ and $\alpha =
    0.05$. The transition from the antiferromagnetic phase (AF) to the
    paramagnetic phase (PA) is second-order at half-filling (thin
    line) and first-order at zero temperature. These two lines merge
    at a tricritical point, $t$. The scales $T_c$ and $\sigma_0$ are
    those defined in Fig.~\ref{f:AFalone-t0dot5}.}
  \label{f:AFalone-t0dot7}
\end{center}
\end{figure}

\begin{figure}[ht]
\begin{center}

\begin{picture}(0,0)%
\includegraphics{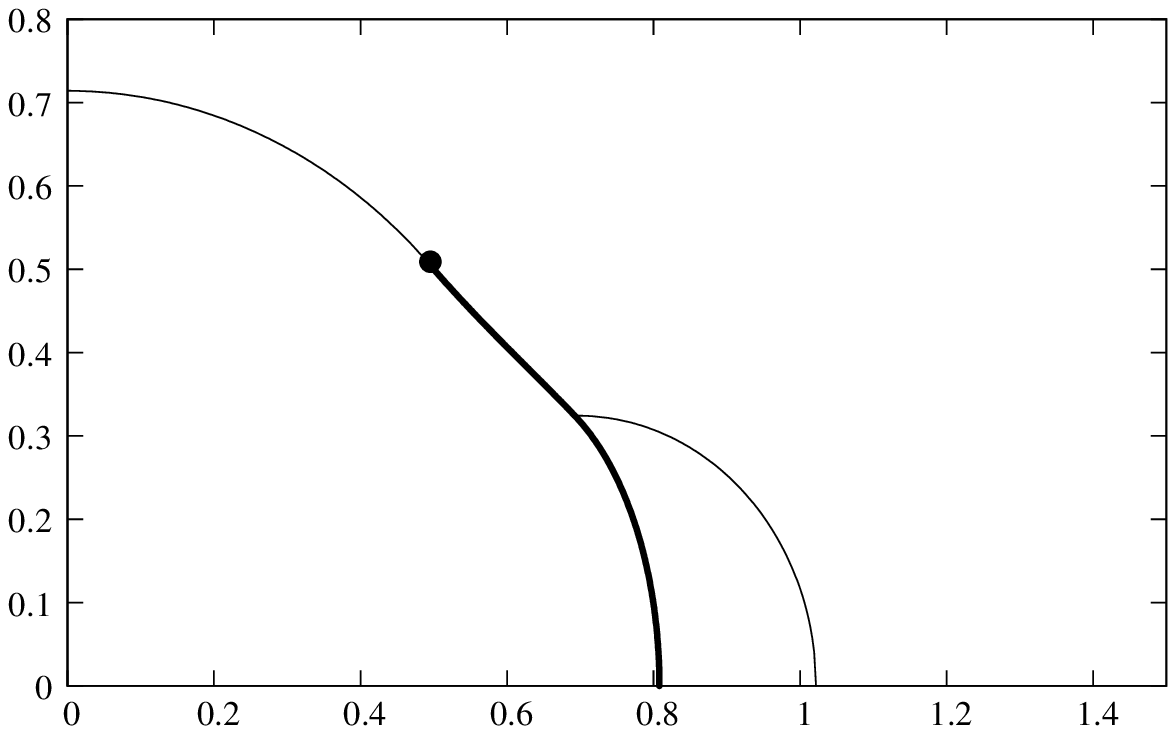}
\end{picture}%
\setlength{\unitlength}{3947sp}%
\begingroup\makeatletter\ifx\SetFigFontNFSS\undefined%
\gdef\SetFigFontNFSS#1#2#3#4#5{%
  \reset@font\fontsize{#1}{#2pt}%
  \fontfamily{#3}\fontseries{#4}\fontshape{#5}%
  \selectfont}%
\fi\endgroup%
\begin{picture}(6178,3787)(822,-4240)
\put(3448,-1571){\makebox(0,0)[lb]{\smash{{\SetFigFontNFSS{12}{14.4}{\familydefault}{\mddefault}{\updefault}{\color[rgb]{0,0,0}$t$}%
}}}}
\put(837,-1981){\makebox(0,0)[lb]{\smash{{\SetFigFontNFSS{12}{14.4}{\familydefault}{\mddefault}{\updefault}{\color[rgb]{0,0,0}$T/T_c$}%
}}}}
\put(3993,-4162){\makebox(0,0)[lb]{\smash{{\SetFigFontNFSS{12}{14.4}{\familydefault}{\mddefault}{\updefault}{\color[rgb]{0,0,0}$\mu/\sigma_0$}%
}}}}
\put(2435,-2526){\makebox(0,0)[lb]{\smash{{\SetFigFontNFSS{12}{14.4}{\familydefault}{\mddefault}{\updefault}{\color[rgb]{0,0,0}AF}%
}}}}
\put(5084,-1815){\makebox(0,0)[lb]{\smash{{\SetFigFontNFSS{12}{14.4}{\familydefault}{\mddefault}{\updefault}{\color[rgb]{0,0,0}PA}%
}}}}
\put(4705,-3246){\makebox(0,0)[lb]{\smash{{\SetFigFontNFSS{12}{14.4}{\familydefault}{\mddefault}{\updefault}{\color[rgb]{0,0,0}SC}%
}}}}
\end{picture}%

  \caption{Phase diagram for $t/t_\mathrm{TH}=0.7$ and $\alpha =
    0.2$. Thin lines are second-order and the thick line is
    first-order; $t$ is a tricritical point. The scales $T_c$ and
    $\sigma_0$ are those defined in Fig.~\ref{f:AFalone-t0dot5}.}
    \label{f:SCemergesout-t0dot7}
\end{center}
\end{figure}

\begin{figure}[ht]
\begin{center}

\begin{picture}(0,0)%
\includegraphics{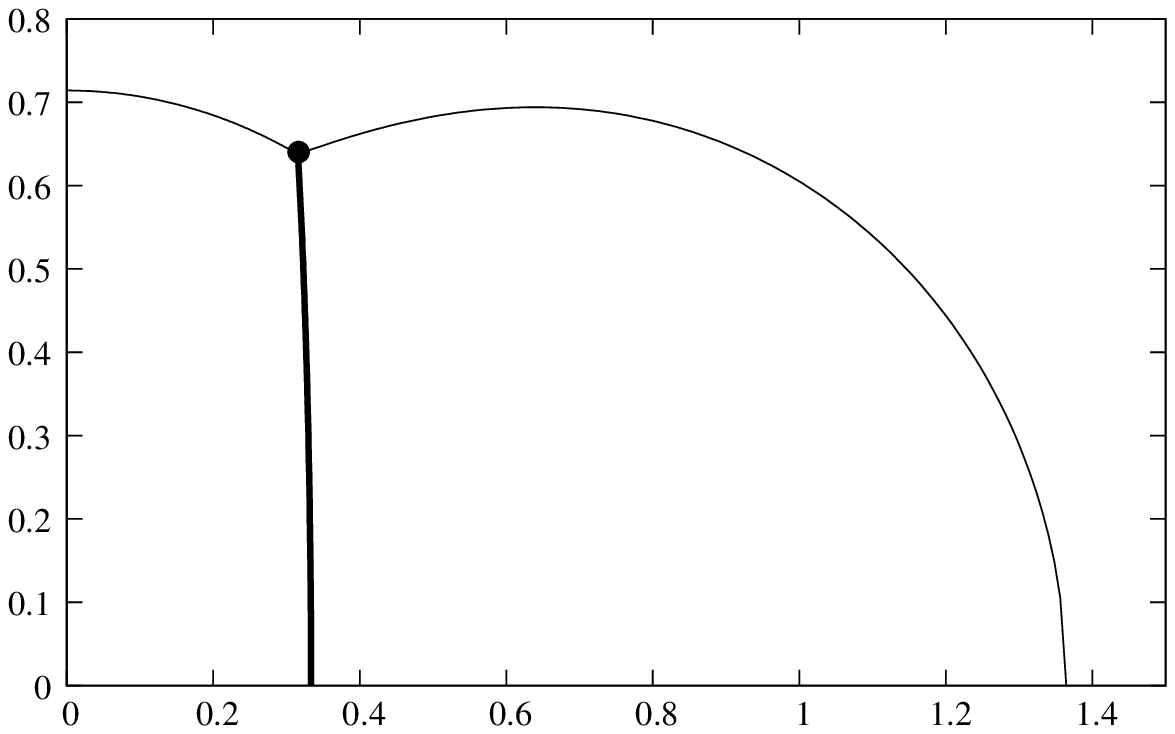}
\end{picture}%
\setlength{\unitlength}{3947sp}%
\begingroup\makeatletter\ifx\SetFigFontNFSS\undefined%
\gdef\SetFigFontNFSS#1#2#3#4#5{%
  \reset@font\fontsize{#1}{#2pt}%
  \fontfamily{#3}\fontseries{#4}\fontshape{#5}%
  \selectfont}%
\fi\endgroup%
\begin{picture}(6125,3827)(875,-4280)
\put(2108,-2283){\makebox(0,0)[lb]{\smash{{\SetFigFontNFSS{12}{14.4}{\familydefault}{\mddefault}{\updefault}{\color[rgb]{0,0,0}AF}%
}}}}
\put(4260,-2293){\makebox(0,0)[lb]{\smash{{\SetFigFontNFSS{12}{14.4}{\familydefault}{\mddefault}{\updefault}{\color[rgb]{0,0,0}SC}%
}}}}
\put(5780,-1085){\makebox(0,0)[lb]{\smash{{\SetFigFontNFSS{12}{14.4}{\familydefault}{\mddefault}{\updefault}{\color[rgb]{0,0,0}PA}%
}}}}
\put(890,-1991){\makebox(0,0)[lb]{\smash{{\SetFigFontNFSS{12}{14.4}{\familydefault}{\mddefault}{\updefault}{\color[rgb]{0,0,0}$T/T_c$}%
}}}}
\put(4163,-4202){\makebox(0,0)[lb]{\smash{{\SetFigFontNFSS{12}{14.4}{\familydefault}{\mddefault}{\updefault}{\color[rgb]{0,0,0}$\mu/\sigma_0$}%
}}}}
\put(2735,-968){\makebox(0,0)[lb]{\smash{{\SetFigFontNFSS{12}{14.4}{\familydefault}{\mddefault}{\updefault}{\color[rgb]{0,0,0}$b$}%
}}}}
\end{picture}%

  \caption{Phase diagram for $t/t_\mathrm{TH}=0.7$ and $\alpha
    =0.8$. Thin lines are second-order and the thick line is
    first-order. $b$ is a bicritical point.  The scales $T_c$ and
    $\sigma_0$ are those defined in Fig.~\ref{f:AFalone-t0dot5}.}
  \label{f:bi-t0dot7}
\end{center}
\end{figure}

{\bf Effect of the hopping term}.  Since the form of the thermodynamic
potential is relatively simple, it is easy to vary the hopping
parameter in order to understand its influence on the phase structure
of the system. Figures~\ref{f:AFalone-t0dot7}---~\ref{f:bi-t0dot7} show
the phase diagrams that are realized by a system with
$t/t_{\mathrm{TH}} = 0.7$ and $\alpha = 0.05$, $0.2$, and $0.8$,
respectively. The resulting phase diagrams exhibit the same basic
topologies found in the previous case.  No additional phase diagrams
are introduced by a larger value of $t$. Increasing the hopping term
has two primary effects. First, the threshold parameters
$\alpha_\Delta$ and $\alpha_b$ are both reduced. Second, larger $t$
tends to increase the region occupied by the superconducting phase and
decrease that occupied by the antiferromagnetic phase. This can be
understood directly from the form of the quasiparticle energies
$\varepsilon_\pm$ in Eq.~(\ref{qp}).  In the antiferromagnetic phase,
$\varepsilon_\pm$ directly depends on the combination
$\sigma^2+t^2$. Increasing $t$ without excessively increasing
$\varepsilon_\pm$ thus requires a reduction in $\sigma$. In the
superconducting phase, however, $\varepsilon_\pm$ varies with $t$ as
$(t\pm\mu)^2+|\Delta|^2$ so that the $\mu$ term tends to reduce the
sensitivity of $\Delta$ with respect to variations of $t$.

{\bf Absence of a mixed-broken symmetry phase}.  In
Sec.~\ref{ss:bicritical}, we found that $\Omega$ has a 
series expansion of the form $\Omega(\sigma,\Delta)
\approx \Omega(0,0) + a_4 |\Delta|^4 + b_4 \sigma^4 + c_4 \sigma^2
|\Delta|^2$ near the bicritical point, where the 
coefficients $a_4$, $b_4$, and $c_4$ are known
functions of $t$ and $\alpha$ which satisfy the inequality $4 a_4 b_4
- c_4^2 < 0$. As a result, a phase with both non-zero $\sigma$ and
non-zero $\Delta$ is never realized. Such a mixed-broken symmetry
state is in fact a solution of the coupled gap equations $\partial
\Omega/\partial \sigma =0$ and $\partial \Omega/\partial \Delta =0$
but is unstable since the matrix of second derivatives has a negative
determinant (i.e., the mixed-broken symmetry phase is necessarily a
saddle point of $\Omega(\sigma,\Delta)$).  In this regard, the present
random matrix result differs from mean-field results for both the
Hubbard\cite{Inui1988,Kyung00} and $t$-$J$ models,\cite{Kyung2002}
for which mixed-broken symmetry phases are found along with a
tetracritical --- rather than a bicritical --- point in the $(\mu,T)$
plane. Coexisting phases have also been reported in a number of
numerical studies including variational cluster perturbation
theory,\cite{Senechal2005} and variational Monte Carlo
methods,\cite{Pathak2008} while other studies report phase
separation.\cite{Heiselberg2008} Calculations from cellular dynamical
mean-field theory show a strong tendency to a homogeneous coexistence
phase at weak coupling and a first-order phase transition at strong
coupling.\cite{Kancharla2007}

The differences arising at mean-field level between the random matrix
approach and microscopic models may well be a consequence of the
neglect of density of states effects in the potential of
Eq.~(\ref{TP-final}). In the absence of a logarithmic divergence in
the gap equations, condensation fields are weakened and the energy
balance between the phases can be upset.  We note, however, that some
authors have questioned whether the phase diagram is necessarily
controlled solely by the Van Hove singularity in the bare density of
states\cite{Pathak2008} as mean-field microscopic results would seem
to show. The discussion of this point shows an example of results that
should not be considered universal or robust; further careful
numerical studies are needed to settle this issue. Although a random
matrix model will not give a definite answer, its comparison to
other mean-field approaches can help in identifying those features
that are sensitive to the specifics of models and their numerical
treatment and which are thus not protected by the symmetries.

{\bf Absence of an exact higher symmetry}. It is interesting to ask
whether the potential of Eq.~(\ref{TP-final}) can exhibit a higher
symmetry, such as the $SO(5)$ symmetry proposed by
Zhang,\cite{Zhang97} either exactly or approximately in certain
regions of the phase diagram. Such a symmetry should manifest itself
in the possibility of writing $\Omega$ as a function of a single
combination of the condensation fields, such as $\sigma^2+|\Delta|^2$.

If we restrict our attention to $\alpha = B/A < 1$ so that AF order
truly competes with SC order, the answer is negative.  Near the
bicritical point, we found that the coefficients in the series
expansion of the thermodynamical potential, $\Omega \approx
\Omega(0,0) + a_4 |\Delta|^4+b_4 \sigma^4 + c_4 |\Delta|^2 \sigma^2$,
satisfy the inequality $4 a_4 b_4 - c_4^2 < 0$. This inequality
eliminates the possibility that the fourth-order terms are a perfect
square. Therefore, no simple symmetry --- allowing for a rotation of
the AF order parameter into the SC order parameter --- can be
identified at the level of two-body correlators. Moreover, we 
have been unable to identify any other symmetry from an expansion 
of $\Omega$ to higher order.

The limit $\alpha \to 1$ can roughly be seen as one of approximate
higher symmetry. In fact, the bicritical point migrates to the
vertical axis, $\mu_b\to 0$, and we have $4 a_4 b_4 - c_4^2 \propto
\mu_b^4 \to 0$. Then, $\Omega$ can be approximated as $\Omega \approx
\Omega(0, 0) + (\sqrt{a_4}\, |\Delta|^2 + \sqrt{b_4} \sigma^2)^2$ with
$a_4 \approx b_4$, a form which shows an approximate higher symmetry.
In the limit $\alpha \to 1$, the antiferromagnetic phase exists only
in the immediate vicinity of half-filling and the superconducting
phase dominants the region of finite chemical potential.  Realizing
such a situation in actual materials would require very specific
relationships between the coupling constants.

\section{Conclusions}
\label{s:C}

In this paper, we have suggested a  mean-field model for
investigating the thermodynamic competition between magnetic and
superconducting orders in a two-dimensional square lattice.  This
model describes interactions at a more microscopic level than the
familiar Hubbard or $t$-$J$ Hamiltonian through the introduction of
density and spin-fluctuation exchanges. The single-particle
Hamiltonian is given a block structure that is dictated by spin,
time-reversal, and bipartite symmetries, and the
detailed dynamics of interactions are replaced by a normal
distribution of random matrix elements. The model is formulated in a
momentum representation so that a coarse description of the first
Brillouin region allows us to introduce a $d$-wave form factor that
possesses the appropriate sign symmetry but neglects detailed momentum
dependences.  Although this approach may seem elaborate at first
sight, the model simplifies dramatically as one proceeds through the
derivation. In fact, the resulting thermodynamic potential has a
simple and well-defined structure that depends solely on the form of
the quasiparticle energies in given condensation fields.

We have explored a number of physically relevant cases for which the
interactions are attractive in both the antiferromagnetic and the
$d$-wave channels and repulsive in the $s$-wave channel.  Such 
interactions naturally place greater weight on spin-fluctuation
exchanges, particularly those involving a momentum transfer
$\sim \mathbf{Q}$. The phase diagram is found to depend 
on a single parameter ratio, $\alpha$, and a limited number of 
topologies appear as a function of $\alpha$. None of these 
topologies allows for a mixed-broken symmetry phase with coexisting
antiferromagnetism and superconductivity, probably as a consequence 
of the absence of the singularity in the bare density of states. Except 
for the  smallest values of $\alpha$, which result in a phase diagram 
involving antiferromagnetism alone, the range of values for $\alpha$ that
correspond to a given topology is rather large. As a result, these
phase topologies should be regarded as robust with respect to moderate
variations of the detailed description of the interactions.

The random matrix approach described here has a broad range of
applications, as we have demonstrated by studying the phase diagrams
of QCD and that of the cuprates. Given the relatively simple structure
that is obtained for the thermodynamic potential, this method is
convenient for obtaining a direct zeroth-order description of the
phase structure. In fact, such a potential could also be written
immediately given only knowledge of the quasiparticle
energies. However, the explicit construction of the microscopic interactions
carries additional and possibly useful information regarding the 
dependence of the coupling
parameters on microscopic variances. In the vicinity of critical
points, the thermodynamical potential can be expanded as a power
series of the condensation fields in a form similar to a
Ginsburg-Landau theory. The information relating microscopic
processes to the coupling parameters thus imposes significant 
constraints on the Ginzburg-Landau coefficients and offers an 
improved understanding of the relation between global properties 
of the system and its microscopic description.

In some sense, this paper can be regarded as ``open source theoretical
physics.''  It contains an ``algorithm'' that can be adapted for use
in other problems.  The shortcoming if this approach is clearly that
it is no more than mean-field theory.  Its merit lies in the fact that
it gives results which are averaged over an ensemble of theories.
Model-dependent details are thus eliminated.  The resulting
thermodynamic potential is dictated by the underlying symmetries of
the problem and is likely to be robust. 

 Here, we have concentrated on the possible topologies of the phase
 diagram and hence on the global minimum of $\Omega$ as a function of
 $\mu$ and $T$. We believe that the results presented here represent a
 generic mean-field phase diagram.  Other features found either in
 theory or experiment are likely to be ``fragile.''  For instance, the
 absence of a mixed-broken symmetry state in the present model
 suggests that this state is sensitive to model-dependent details,
 numerical approximations, or to the detailed properties of the sample
 studied.  Indeed, the mixed-broken symmetry state is not generic in
 the cuprates.\cite{Kancharla2007} More comprehensive studies of
 $\Omega$ can be useful in revealing, e.g., the competition between
 local minima in $\Omega$ and other relatively fragile structures that
 could be sensitive to model-dependent details and numerical
 approximations.  In this sense, we suggest that the present methods
 may provide a useful complement to the investigation of detailed
 models of these systems.

\appendix

\section{Calculation of the thermodynamic potential}
\label{a:calculations-TP}

In this appendix, we sketch the calculations leading to the
thermodynamic potential, $\Omega$, of Eq.~(\ref{TP-final}). The
calculations closely follow those of the random matrix model 
for QCD given in Ref.\cite{VanJac00}.

We start with the single-fermion Hamiltonian of Eq.~(\ref{Hint}) and
integrate $\exp(-\psi^{\dagger}H_\mathrm{int}\psi)$ over the matrix
elements of $H_\mathrm{int}$ according to the normal probability
distribution $P(H_\mathrm{int})$ of Eq.~(\ref{PH-final}). This
integration can be written as
\begin{eqnarray}
\int DH_{\mathrm{int}} P(H_\mathrm{int}) \,e^{-\psi^\dagger H_{\mathrm{int}}\psi} =
e^{Y},
\label{Y}
\end{eqnarray}
where the four-fermion potential, $Y$, receives a contribution from each
of the eight independent block matrices
\begin{eqnarray}
  Y = Y_{B0d} + Y_{\mathbf{B}_d} + Y_{B0o} + Y_{\mathbf{B}_o} + Y_{C0d} + 
Y_{\mathbf{C}_d} + Y_{C0o} + Y_{\mathbf{C}_o}.
\end{eqnarray}
An integration over the matrix elements describing density fluctuations gives
\begin{eqnarray}  
Y_{B0d} & = & {1 \over 32 M \Sigma^2_{B0d}} \sum_{ij}\left(
1_i\:1_{j}+2_i\:2_{j}+3_j\:3_{i}+4_j\:4_{i}\right)\left(\mathrm{h.c.}\right),
\label{YB0d}\\
  Y_{B0o} & = & {1 \over 32 M \Sigma^2_{B0o}} \sum_{ij}\left(
1_i\:2_{j}+2_i\:1_j + 3_j\:4_i+ 4_j\:3_{i} \right)\left(\mathrm{h.c.}\right), 
\label{YB0o}\\
  Y_{C0d} & = & {1 \over 32 M \Sigma^2_{C0d}} \sum_{ij}\left(
1_i\:3_{j}+2_i\:4_{j}+1_j\:3_{i}+2_j\:4_{i}\right)
\left(\mathrm{h.c.}\right),
\label{YC0d}\\
  Y_{C0o} & = & {1 \over 32 M \Sigma^2_{C0o}} \sum_{ij}
\left(1_i\:4_{j}+2_i\:3_{j}+1_j\:4_{i}+2_j\:3_{i}\right)\left(\mathrm{h.c.}\right),
\label{YC0o}
\end{eqnarray}
where $(\mathrm{h.c.})$ indicates Hermitian conjugation and 
$1_i\:1_{j}$ is a compact notation for 
$\sum_{\alpha}\psi^\dagger_{1i\alpha}\psi_{1j\alpha}$.
(Here, $\psi_{1j\alpha}$ represents a state with the momentum label $j$ in
region $1$ and a spin index $\alpha$.)  Integration over the matrix 
elements of the blocks describing spin fluctuations produces the terms
\begin{eqnarray}
  Y_{\mathbf{B}_d} & = & {1 \over 32 M \Sigma^2_{\mathbf{B}_d}} \sum_{ij}\left(
      1_i \,\mbox{\boldmath $\sigma$}\, 1_{j}+
      2_i \,\mbox{\boldmath $\sigma$}\,2_{j}+
      3_j \,\mbox{\boldmath $\sigma$}\,3_{i} + 
      4_j\,\mbox{\boldmath $\sigma$}\,4_{i}\right)
      \cdot
      \left(\mathrm{h.c.}\right),
\label{YBvecd}\\
  Y_{\mathbf{B}_o} & = & {1 \over 32 M \Sigma^2_{\mathbf{B}_o}} \sum_{ij}\left(
      1_i \,\mbox{\boldmath $\sigma$}\,2_{j} + 
      2_i \,\mbox{\boldmath $\sigma$}\,1_{j} + 
      3_j \,\mbox{\boldmath $\sigma$}\,4_{i} + 
      4_j \,\mbox{\boldmath $\sigma$}\,3_{i} \right) 
      \cdot
      \left(\mathrm{h.c.}\right),
\label{YBveco}\\
  Y_{\mathbf{C}_d} & = & {1 \over 32 M \Sigma^2_{\mathbf{C}_d}} \sum_{ij}\left(
      1_i\,\mbox{\boldmath $\sigma$}\,3_{j} +
      2_i\,\mbox{\boldmath $\sigma$}\,4_{j} +
      1_j\,\mbox{\boldmath $\sigma$}\,3_{i} + 
      2_j\,\mbox{\boldmath $\sigma$}\,4_{i}\right)
     \cdot \left(\mathrm{h.c.}\right), 
\label{Cvecd}\\
  Y_{\mathbf{C}_o} & = & {1 \over 32 M \Sigma^2_{\mathbf{C}_o}}
  \sum_{ij}\left(
  1_i\,\mbox{\boldmath$\sigma$}\,4_{j}+2_i\,\mbox{\boldmath
    $\sigma$}\,3_{j} + 1_j\,\mbox{\boldmath
    $\sigma$}\,4_{i}+2_j\,\mbox{\boldmath $\sigma$}\,3_{i}
  \right)\cdot \left(\mathrm{h.c.}\right),
\label{YCveco}
\end{eqnarray}
where the notation $1_i\,\mbox{\boldmath $\sigma$}\,2_{j}$ stands
for $\sum_{\alpha\beta}\psi^\dagger_{1i\alpha}
\mbox{\boldmath $\sigma$}_{\alpha\beta} \psi_{2j\beta}$.

Next, fermion fields are rearranged to make  the condensation
channels apparent. Schematically, terms of the form $\psi^\dagger_i
\psi_j^{\phantom{\dagger}} \psi^\dagger_j \psi_i^{\phantom{\dagger}}$
(with the sum over $i$ and $j$ implied) are brought into the form $-
\psi^\dagger_i \psi_i^{\phantom{\dagger}} \psi^\dagger_j
\psi_j^{\phantom{\dagger}}$, which gives rise to condensates
$\sim\psi^\dagger_i\psi_i^{\phantom{\dagger}}$ that are relevant for
antiferromagnetism. Similarly, terms of the form $\psi^\dagger_i
\psi_j^{\phantom{\dagger}} \psi^\dagger_i \psi_j^{\phantom{\dagger}}$
are rewritten as $\psi^\dagger_i \psi^\dagger_i
\psi_j^{\phantom{\dagger}} \psi_j^{\phantom{\dagger}}$, which contain
condensates of the form $\sim
\psi_i^{\phantom{\dagger}}\psi_i^{\phantom{\dagger}}$ and are relevant
for superconductivity.  Uncrossing necessitates keeping track of
momentum and spin indices.  To uncross spin indices, we use the
$SU(2)$ Fierz identities
\begin{eqnarray}
  \delta_{ab}\,\delta_{cd} & = & {1 \over 2}\,\delta_{ad}\,\delta_{cb}
  + {1\over 2}\,\mbox{\boldmath $\sigma$}_{ad}\cdot \mbox{\boldmath
    $\sigma$}_{cb},\\ \mbox{\boldmath
    $\sigma$}_{ab}\cdot\mbox{\boldmath $\sigma$}_{cd} & = & {3\over
    2}\,\delta_{ad}\delta_{cb} - {1\over 2}\,\mbox{\boldmath
    $\sigma$}_{ad}\cdot\mbox{\boldmath $\sigma$}_{cb}
\end{eqnarray}
for the antiferromagnetism channel and the identities
\begin{eqnarray}
  \delta_{ab}\,\delta_{cd} & = & {1 \over 2}\,\delta_{ac}\,\delta_{db}
  + {1\over 2}\,\mbox{\boldmath $\sigma$}_{ac}\cdot\mbox{\boldmath
    $\sigma$}_{db},\\ \mbox{\boldmath $\sigma$}_{ab}\cdot
    \mbox{\boldmath $\sigma$}_{cd} & = &
  {1\over 2}\,\delta_{ac}\delta_{db} + {1\over
    2}\,(\sigma_1)_{ac}(\sigma_1)_{db} - {3\over
    2}\,(\sigma_2)_{ac}(\sigma_2)_{db} + {1\over
    2}\,(\sigma_3)_{ac}(\sigma_3)_{db}
\end{eqnarray}
for superconductivity channels. To uncross momentum indices, each
quadratic term $\psi^\dagger_r \psi_s^{\phantom{\dagger}}$ (where 
$r$ and $s$ now denote momentum indices) is written as an element of a
$4\times 4$ momentum matrix which is then decomposed on a complete
basis of $16$ Hermitian Dirac matrices\cite{VanJac00} that satisfy
\begin{eqnarray}
  \mathrm{Tr}(\Gamma_k \Gamma_l) & = & 4 \delta_{kl}\quad (k,l=1,\ldots,16).
\end{eqnarray}
There exists a representation in which this basis contains the
operators $\Gamma_{AF}$, $\Gamma_{SC-d}$ and $\Gamma_{SC-s}$, which
were introduced, respectively, in Eqs.~(\ref{GammaAF}),
(\ref{GammaSC-d}), and (\ref{GammaSC-s}), as well as an operator that
is proportional to the hopping term, $\Gamma_t$, of
Eq.~(\ref{Gammat}). With this complete basis defined, we uncross
momentum according to the Fierz identities
\begin{eqnarray}
  (\Gamma_k)_{ab}(\Gamma_l)_{cd} & = & \sum_{mn} x_{klmn} (\Gamma_m)_{ad}
(\Gamma_n)_{cb},\\
 x_{klmn} & = & {1\over 16}\,\mathrm{Tr}
             \left(\Gamma_k\Gamma_n\Gamma_l\Gamma_m\right)
\end{eqnarray}
for terms relevant for antiferromagnetism and
\begin{eqnarray}
  (\Gamma_k)_{ab}(\Gamma_l)_{cd} & = & \sum_{mn} x_{klmn} (\Gamma_m)_{ac}
(\Gamma_n)_{db},\\
 x_{klmn} & = & {1\over 16}\,\mathrm{Tr}
             \left(\Gamma_k\Gamma_n\Gamma_l^T\Gamma_m\right)
\end{eqnarray}
for terms relevant for superconductivity. Here, $\mathrm{Tr}$ denotes
a trace and $\Gamma^T$ is the transpose of $\Gamma$.

Restricting ourselves to the antiferromagnetic and superconducting
channels, uncrossing gives us the four-fermion interaction
\begin{eqnarray}
  Y & = & Y_{AF} + Y_{SC-d} + \ldots \label{Y1}\\ 
  &=&  a_{AF} \,\left(\psi^\dagger
  \Gamma_{AF}\sigma_3\psi\right)^2 + 
  a_{SC-s}\,\left(\psi^\dagger\Gamma_{SC-s}\sigma_2\psi^\dagger\right) 
  \left(\psi^{\phantom{\dagger}}\Gamma_{SC-s}\sigma_2
   \psi^{\phantom{\dagger}}\right) \nonumber \\
  && +
  a_{SC-d}\,\left(\psi^\dagger\Gamma_{SC-d}\sigma_2\psi^\dagger\right) 
  \left(\psi^{\phantom{\dagger}}\Gamma_{SC-d}\sigma_2\psi^{\phantom{\dagger}}
\right),
\label{Y2}
\end{eqnarray}
where a summation over spin, momentum, and random matrix indices is
implied in each fermion bilinear.  The coefficients in 
the respective channels are given by
\begin{eqnarray}
  a_{AF} & = & {1\over 256 M} \left(- {1\over \Sigma_{B0d}^2} - 
{1\over \Sigma_{B0o}^2} - {1\over \Sigma_{C0d}^2} - {1\over \Sigma_{C0o}^2} + 
{1\over \Sigma_{\mathbf{B}_d}^2} + {1\over \Sigma_{\mathbf{B}_o}^2} +
{1\over \Sigma_{\mathbf{C}_d}^2} + {1\over \Sigma_{\mathbf{C}_o}^2}
\right),\label{aAF}\\
  a_{SC-s} & = & {1\over 256 M} \left({1\over \Sigma_{B0d}^2} + 
{1\over \Sigma_{B0o}^2} + {1\over \Sigma_{C0d}^2} + 
{1\over \Sigma_{C0o}^2} 
- {3\over \Sigma_{\mathbf{B}_d}^2} - {3\over \Sigma_{\mathbf{B}_o}^2}
- {3\over \Sigma_{\mathbf{C}_d}^2}- {3\over \Sigma_{\mathbf{C}_o}^2}\right),\\
  a_{SC-d} & = & {1\over 256 M} \left({1\over \Sigma_{B0d}^2} -
{1\over \Sigma_{B0o}^2} + {1\over \Sigma_{C0d}^2} - 
{1\over \Sigma_{C0o}^2} 
- {3\over \Sigma_{\mathbf{B}_d}^2} + {3\over \Sigma_{\mathbf{B}_o}^2}
- {3\over \Sigma_{\mathbf{C}_d}^2} + {3\over \Sigma_{\mathbf{C}_o}^2}\right).
\label{aSCd}
\end{eqnarray}
Here, a positive (negative) coefficient corresponds to an attractive
(repulsive) channel. From the above equations, we thus see that
interactions for which the inverse variances $\Sigma^2_{\mathbf{B}_d},
\Sigma^2_{\mathbf{B}_o}$, $\Sigma^2_{\mathbf{C}_d}$, or
$\Sigma^2_{\mathbf{C}_o}$ are small compared to
$\Sigma^2_{B0d},\Sigma^2_{B0o}$, $\Sigma^2_{C0d}$, or $\Sigma^2_{C0o}$
are attractive in the antiferromagnetic channel. Such interactions
favor the exchange of spin fluctuations over the exchange of density
fluctuations. Similarly, interactions can be made attractive in the
$d$-wave channel and repulsive in the $s$-wave channel by favoring the
elements of the off-diagonal block matrices $\mathbf{B}_o$ and
$\mathbf{C}_o$ over those of the diagonal blocks $\mathbf{B}_d$ and
$\mathbf{C}_d$. Such interactions favor the exchange of spin fluctuations
with a large momentum transfer, $\sim\mathbf{Q}$.

Below, we will assume that the random interactions favor
antiferromagnetism and superconductivity in the $d$-wave channel and
are repulsive for $s$-wave pairs; this situation corresponds to a
particular choice for the variances such that $a_{AF} >0$ and 
$a_{SC-d}>0$ whereas $a_{SC-s} < 0$.  We will thus neglect the
$s$-wave channel in the remainder of the calculations.  In this case,
combining Eqs.~(\ref{partition-function}) and (\ref{Y}) yields a
partition function of the form
\begin{eqnarray}
  Z(\mu,T) = \int {\cal D}\psi^\dagger {\cal D}\psi^{\phantom{\dagger}}
e^{-\psi^\dagger H_0 \psi + Y},
\end{eqnarray}
where, according to Eqs.~(\ref{Y1}) and (\ref{Y2}),  
\begin{eqnarray}
  Y \sim a_{AF} \,\left(\psi^\dagger \Gamma_{AF}\sigma_3\psi\right)^2 +
      a_{SC-d}\,\left(\psi^\dagger\Gamma_{SC-d}\sigma_2\psi^\dagger\right) 
  \left(\psi^{\phantom{\dagger}}\Gamma_{SC-d}\sigma_2\psi^{\phantom{\dagger}}
\right).
\label{Y-before-HS}
\end{eqnarray}
The quartic fermion terms can now be written as the difference of two
squares. Each square is linearized by the use of a
Hubbard-Stratonovitch transformation,
\begin{eqnarray}
  e^{A Q^2} \sim \int dx \exp\left(-  {x^2\over 4 A} - Q
  x\right),
\end{eqnarray}
which introduces an auxiliary field $x$. When applied to
Eq.~(\ref{Y-before-HS}), such transformations introduce a real field,
$\sigma$, to be associated with antiferromagnetism and a complex
field, $\Delta$, to be related to superconductivity. The partition
function is then written as
\begin{eqnarray}
  Z(\mu,T) & \sim & \int d\sigma d\Delta d\Delta^* \int {\cal D}\psi^\dagger
{\cal D}\psi^{\phantom{\dagger}} 
\exp\left(- \psi^\dagger H_0 \psi - {|\Delta|^2 \over 4 a_{SC-d}} -
{\sigma^2 \over 4 a_{AF}}\right) 
\nonumber\\
  & & \times \exp\left( - \sigma \psi^\dagger \Gamma_{AF} \sigma_3 \psi 
 - {\Delta\over 2} \psi^\dagger \Gamma_{SC-d} \sigma_2 \psi^\dagger 
 - {\Delta^*\over 2} \psi^{\phantom{\dagger}} \Gamma_{SC-d} \sigma_2 
\psi^{\phantom{\dagger}} \right),
\label{pf2}
\end{eqnarray}
where $H_0 = -\mu +\Omega_T +\Gamma_t$ [see Eq.~(\ref{H0})]. Using the
spinor $\Psi = (\psi_\uparrow , \psi^\dagger_\downarrow)$, the fermion
bilinears can be arranged in the Gorgov form $\Psi^\dagger \tilde{H}\Psi$,
with
\begin{eqnarray}
  \tilde{H} = \left(
  \begin{array}{cc}
    -\mu + \Omega_T + \Gamma_t  + \sigma \Gamma_{AF}  &
    - i \Delta \Gamma_{SC-d}
    \\
    i \Delta^* \Gamma_{SC-d}
    & 

    \mu - \Omega_T - \Gamma_t + \sigma \Gamma_{AF}

    \\
  \end{array}
              \right)
\label{tildeH}
\end{eqnarray}
Then, integrating over the fermion fields gives 
\begin{eqnarray}
  \int {\cal D}\psi^\dagger {\cal D}\psi e^{- \Psi^\dagger \tilde{H} \Psi} =
\mathrm{Det}[\tilde{H}],
\end{eqnarray}
so that the partition function in Eq.~(\ref{pf2}) can be written as
\begin{eqnarray}
  Z(\mu,T)  = \int d\sigma d\Delta d\Delta^* e^{- 8 M \Omega(\sigma,\Delta)},
\end{eqnarray}
where $\Omega$ is the thermodynamic potential 
\begin{eqnarray}
  \Omega & = & {1 \over 32 M} \left( {|\Delta|^2 \over  a_{SC-d}} +
{\sigma^2 \over  a_{AF}} - 4 \log\mathrm{Det}(\tilde{H})\right). 
\end{eqnarray}
Calculating the determinant of $\tilde{H}$ in Eq.~(\ref{tildeH}) then gives
\begin{eqnarray}
  \Omega(\sigma,\Delta) & = & A |\Delta|^2 + B \sigma^2 - {1\over 4}\,
\log((\sqrt{\sigma^2+t^2}-\mu)^2+|\Delta|^2+ T^2) \nonumber\\
&& - {1\over 4}\,
\log((\sqrt{\sigma^2+t^2}+\mu)^2+|\Delta|^2+ T^2),
\label{Omega1}
\end{eqnarray}
with 
\begin{eqnarray}
  A & = & {8} \left({1\over \Sigma_{B0d}^2} -
{1\over \Sigma_{B0o}^2} + {1\over \Sigma_{C0d}^2} - 
{1\over \Sigma_{C0o}^2} 
- {3\over \Sigma_{\mathbf{B}_d}^2} + {3\over \Sigma_{\mathbf{B}_o}^2}
- {3\over \Sigma_{\mathbf{C}_d}^2} + {3\over \Sigma_{\mathbf{C}_o}^2}\right)^{-1},
\label{A}\\
 B & = & 8 \left(- {1\over \Sigma_{B0d}^2} - {1\over \Sigma_{B0o}^2}
 - {1\over \Sigma_{C0d}^2} -  {1\over \Sigma_{C0o}^2} + 
{1\over \Sigma_{\mathbf{B}_d}^2} + {1\over \Sigma_{\mathbf{B}_o}^2} +
{1\over \Sigma_{\mathbf{C}_d}^2} + {1\over \Sigma_{\mathbf{C}_o}^2}\right)^{-1}\label{B}.
\end{eqnarray}
We can evaluate the properties of the competing phases by determining
the minima of the potential $\Omega$. In the thermodynamic limit,
where $M$ is taken to infinity, these minima will give the exact
solutions for the system since a saddle-point evaluation gives
$\lim_{M\to \infty} (1/8M) \mathrm{ln}Z = -
\mathrm{min}_{\sigma,\Delta}(\Omega)$.

We mentioned earlier an alternative to the bipartite
symmetry. This alternative choice consists in taking the matrix
elements between states $\mathbf{p}_1+\mathbf{Q}$ and
$\mathbf{p}_2+\mathbf{Q}$ equal to those between $\mathbf{p}_1$ and
$\mathbf{p}_2$, so that Eq.~(\ref{bipartite-block-structure}) becomes
 \begin{eqnarray}
  H_{\mu} = \left(
                      \begin{array}{cc}
                        D_\mu & E_\mu \\
                        E_\mu^\dagger & D_\mu \\
                      \end{array}
                    \right),
\label{bipartite-block-structure2}
\end{eqnarray}
where now $E_\mu$ are complex. Such a choice modifies the four-fermion
potentials $Y_{B0o}$, $Y_{C0o}$,$Y_{\mathbf{B}_{o}}$, and $Y_{\mathbf{C}_o}$ as
\begin{eqnarray}
  Y_{B0o} & = & {1 \over 8 M \Sigma^2_{B0o}} \sum_{ij}\left(
1_i\:2_{j}+ 4_j\:3_{i} \right)\left(\mathrm{h.c.}\right), 
\label{YB0o2}\\
  Y_{C0o} & = & {1 \over 8 M \Sigma^2_{C0o}} \sum_{ij}
\left(1_i\:4_{j}+2_j\:3_{i}\right)\left(\mathrm{h.c.}\right),
\label{YC0o2}\\
  Y_{\mathbf{B}_o} & = & {1 \over 8 M \Sigma^2_{\mathbf{B}_o}} \sum_{ij}\left(
      1_i \,\mbox{\boldmath $\sigma$}\,2_{j}+ 
      4_j \,\mbox{\boldmath $\sigma$}\,3_{i} \right) 
      \cdot
      \left(\mathrm{h.c.}\right),
\label{YBveco2}\\
  Y_{\mathbf{C}_o} & = & {1 \over 8 M \Sigma^2_{\mathbf{C}_o}} \sum_{ij}\left(
     1_i\,\mbox{\boldmath $\sigma$}\,4_{j}+
    2_j\,\mbox{\boldmath $\sigma$}\,3_{i}\right)\cdot
     \left(\mathrm{h.c.}\right).
\label{YCveco2}
\end{eqnarray}
The  coupling constants $a_{AF}$, $a_{SC-s}$, and $a_{SC-d}$ become
\begin{eqnarray}
  a_{AF} & = & {1\over 256 M} \left(- {1\over \Sigma_{B0d}^2} - 
{1\over \Sigma_{C0d}^2} + {1\over \Sigma_{\mathbf{B}_d}^2} +
{1\over \Sigma_{\mathbf{C}_d}^2}\right),\label{aAF2}\\
  a_{SC-s} & = & {1\over 256 M} \left({1\over \Sigma_{B0d}^2} + 
{2\over \Sigma_{B0o}^2} + {1\over \Sigma_{C0d}^2} + 
{2\over \Sigma_{C0o}^2} 
- {3\over \Sigma_{\mathbf{B}_d}^2} - {6\over \Sigma_{\mathbf{B}_o}^2}
- {3\over \Sigma_{\mathbf{C}_d}^2}- {6\over \Sigma_{\mathbf{C}_o}^2}\right),\\
  a_{SC-d} & = & {1\over 256 M} \left({1\over \Sigma_{B0d}^2} -
{2\over \Sigma_{B0o}^2} + {1\over \Sigma_{C0d}^2} - 
{2\over \Sigma_{C0o}^2} 
- {3\over \Sigma_{\mathbf{B}_d}^2} + {6\over \Sigma_{\mathbf{B}_o}^2}
- {3\over \Sigma_{\mathbf{C}_d}^2} + {6\over \Sigma_{\mathbf{C}_o}^2}\right).
\label{aSCd2}
\end{eqnarray}
Again, the interaction can be made attractive in the AF channel by
favoring spin over density exchanges, while it its attractive in the
$d$-wave channel and repulsive for $s$-wave pairing when large
momentum transfers are favored. The resulting thermodynamic 
potential has the form of Eq.~(\ref{Omega1}), although with slightly
different expressions for $A$ and $B$. The main results in the text
remain valid with this alternative choice.


\end{document}